\documentclass[aps, prd, superscriptaddress, nofootinbib, preprintnumbers]{revtex4}
\usepackage{amsmath}
\usepackage{graphicx}
\usepackage{color}

\newcommand{\gtrsim}{\mathrel{\mathpalette\vereq>}}

\newcommand{\chushi}[1]{}

\begin{document}

\title{Discovering walking technirho mesons at the LHC}     
\author{Masafumi Kurachi} \thanks{\tt kurachi@kmi.nagoya-u.ac.jp}
      \affiliation{ Kobayashi-Maskawa Institute for the Origin of Particles and the Universe (KMI) \\ 
Nagoya University, Nagoya 464-8602, Japan.}
\author{Shinya Matsuzaki}\thanks{\tt synya@hken.phys.nagoya-u.ac.jp}
      \affiliation{ Institute for Advanced Research, Nagoya University, Nagoya 464-8602, Japan.}
      \affiliation{ Department of Physics, Nagoya University, Nagoya 464-8602, Japan.}
\author{{Koichi Yamawaki}} \thanks{
      {\tt yamawaki@kmi.nagoya-u.ac.jp}}
      \affiliation{ Kobayashi-Maskawa Institute for the Origin of Particles and the Universe (KMI) \\ 
 Nagoya University, Nagoya 464-8602, Japan.}

\begin{abstract}
We formulate a scale-invariant hidden local symmetry (HLS) 
as a low-energy effective theory of the walking technicolor (WTC) 
which includes the technidilaton, technipions, and technirho mesons as the low-lying spectra.   
As a benchmark for LHC phenomenology, 
we in particular focus on the one-family model of WTC having eight technifermion flavors, 
which can be at energy scales relevant to the reach of the LHC described by the scale-invariant HLS based on the manifold 
$[SU(8)_L \times SU(8)_R]_{\rm global} \times SU(8)_{\rm local}/SU(8)_V$,    
where $SU(8)_{\rm local}$ is the HLS and the global $SU(8)_L \times SU(8)_R$ symmetry is partially 
gauged by $SU(3) \times SU(2)_L \times U(1)_Y$ of the standard model.     
Based on the scale-invariant HLS, we evaluate the coupling properties of the technirho mesons 
and place limits on the masses from the current LHC data. 
Then, implications for future LHC phenomenology are discussed by focusing on 
the technirho mesons produced through the Drell-Yan process.  
We find that the color-octet technirho decaying to the technidilaton along with the gluon 
is of interest as the discovery channel at the LHC, which would provide  a characteristic 
signature to probe the one-family WTC.  
\end{abstract}
\maketitle

\section{Introduction}
The Higgs boson with a mass of 125 GeV was discovered at the LHC. 
However, the dynamical origin of electroweak (EW) symmetry breaking 
and of the Higgs are still mysterious and would be explained by the physics beyond the standard model (SM). 
Technicolor (TC)~\cite{Weinberg:1975gm,Susskind:1978ms,Farhi:1980xs} is a well-motivated model for the dynamical origin of EW symmetry breaking 
in a way similar to the established mechanism in QCD which breaks  
the chiral symmetry (and hence the EW symmetry as well) dynamically via the fermion pair condensate. 
However, the original TC was ruled out a long time ago by the notorious flavor-changing neutral currents (FCNC) problem, 
and more dramatically by the recent discovery of the 125 GeV Higgs which cannot be accounted for by the TC dynamics of a simple QCD scale-up.  

Fortunately, both problems are simultaneously solved by 
walking technicolor (WTC)~\cite{Yamawaki:1985zg,Yamawaki:1996vr}, 
which was proposed based on the scale-symmetric dynamics of 
the ladder Schwinger-Dyson (SD) equation: 
with the scale symmetry, WTC predicted
 a large anomalous dimension $\gamma_m = 1$ as a solution to the FCNC problem,\footnote{
A similar solution to the FCNC problem was given without the notion of 
the scale symmetry/technidilaton and the anomalous dimension~\cite{Holdom:1984sk}.}  
and at the same time predicted
 a light composite Higgs -- known as a ``technidilaton'' (TD)~\cite{Yamawaki:1985zg, Bando:1986bg} --
that is a pseudo-Nambu-Goldstone (NG) boson of the scale symmetry  broken spontaneously 
(and also explicitly) by the technifermion condensate. 
It was shown that the 
TD can account for the 125 GeV Higgs, with couplings that are consistent with the current LHC data of 
the 125 GeV Higgs (see below)~\cite{Matsuzaki:2012gd,Matsuzaki:2012vc,Matsuzaki:2012mk,Matsuzaki:2012xx}.  
Thus the origin of the Higgs mass is dynamically explained by the  scale of the chiral condensate in WTC.

The mass of the TD as a pseudo-NG boson comes from the nonperturbative trace anomaly due to the chiral condensate and can be estimated 
through the partially conserved dilatation current (PCDC) relation~\cite{Yamawaki:1985zg, Bando:1986bg}. 
A precise ladder evaluation of $m_\phi F_\phi$ based on this PCDC relation reads~\cite{Hashimoto:2010nw}: 
$(m_\phi F_\phi)^2 \simeq  0.154 \cdot N_f N_c\cdot m_D^4 \simeq \left(2.5 \cdot v_{\rm EW}^2\right)^2 \cdot \left[(8/N_f)(4/N_c)\right]$, where $v_{\rm EW}^2=(246\,  {\rm GeV})^2 
= N_D F_\pi^2 \simeq 0.028 \cdot N_f N_c \cdot m_D^2$ (Pagels-Stokar formula),
with $N_D (=N_f/2)$ being the number of the electroweak doublets. 
Note the scaling $m_\phi/v_{\rm EW} \sim
1/\sqrt{N_f N_c}$,  which implies  that a light TD $m_\phi/v_{\rm EW} \ll 1$ is naturally realized for $N_f \gg 1$ (as well as $N_c\gg 1$) 
as in the large-$N_f$ QCD.~\footnote{ 
Thus the mass of the LHC Higgs,  $m_\phi \simeq
125\, {\rm GeV} \simeq v_{\rm EW}/2$, can be obtained~\cite{Matsuzaki:2012gd}, when we take $v_{\rm EW}/F_\phi 
=2F_\pi/F_\phi \simeq 1/5=0.2$  ($v_{\rm EW}= 2 F_\pi$ for $N_c=4, N_f=8$ in the one-family model, see below).  
Amazingly, this value of $F_\phi$ turned out to be
consistent with the LHC Higgs data (best fit $v_{\rm EW}/F_\phi \simeq 0.22$)~\cite{Matsuzaki:2012mk}
}~\footnote{
One might think that such a large $N_f$ (and $N_c$) would result in the so-called $S$ parameter problem~\cite{Peskin:1990zt+}.  
The large $S$ from the TC sector, however,  is not necessarily in conflict with the experimental
value of the $S$ from the precision EW measurements, since the contributions from the TC sector 
can easily be cancelled by strong mixing with the SM fermion contribution
through the extended TC (ETC) interactions, as in the fermion delocalization of the Higgsless model~\cite{Cacciapaglia:2004rb}. 
 Moreover,   even within the TC sector alone, 
 there exists a way to resolve this problem as demonstrated in the holographic model, where 
 where we can reduce  $S \sim 4\pi (N_D F_\pi^2)/M_\rho^2 = 4\pi v_{\rm EW}^2/M_\rho^2$ by tuning the holographic parameter 
(roughly corresponding to increasing the technirho mass $M_\rho$), in a way 
consistent with the TD mass of 125 GeV and all the current LHC data
for the 125 GeV Higgs~\cite{Matsuzaki:2012xx}. 
}

More recently, in another approach using holographic WTC~\cite{Haba:2010hu} 
it was  shown~\cite{Matsuzaki:2012xx} that the strong gluon dynamics via the large technigluon condensate  can realize a parametrically massless TD limit $m_\phi/v_{\rm EW}  \rightarrow 0+$ and hence naturally realize $m_\phi \simeq v_{\rm EW}/2  \simeq 125$ GeV, 
consistent with the LHC Higgs data. 
Similar arguments for realizing a parametrically light dilaton were given in somewhat different contexts~\cite{Elander:2009pk}.

Amazingly, the recent lattice results~\cite{Aoki:2013xza} in fact indicate that the $SU(3)$ gauge theory with eight fundamental fermions 
($N_f=8$ QCD) possesses a walking nature, with the anomalous dimension $\gamma_m \simeq 1$.
Most remarkably, it has been shown 
in the lattice  $N_f=8$ QCD~\cite{Aoki:2014oha} that  
there in fact  exists a flavor-singlet scalar meson that is as light as the pions 
for a small fermion-mass region, which thus can be a composite Higgs (in the form of the TD) in the chiral limit.

Thus, special interest in the context of lattice studies has recently been paid to the 
one-family model (or Farhi-Susskind model)~\cite{Farhi:1979zx,Farhi:1980xs} as  a candidate theory for the WTC. 
The model is the most straightforward version of the 
extended TC (ETC) model~\cite{Dimopoulos:1979es} 
which incorporates the mechanism of producing masses for the SM fermions. 
The one-family model consists of $N_{\rm TC}$ copies of a whole generation of the SM fermions;  
therefore, the TC sector of the model is a $SU(N_{\rm TC})$ gauge theory with eight fundamental Dirac fermions, $N_f=8$ 
(i.e., four weak doublets, $N_D=N_f/2=4$, with the NG boson decay constant $F_\pi \simeq 246\, {\rm GeV}/\sqrt{N_D} = 123$ GeV).

The global chiral symmetry-breaking pattern of the one-family model is 
$G/H= SU(8)_L \times SU(8)_R/SU(8)_V$,  
which is described by the usual nonlinear chiral Lagrangian based on the manifold $G/H$ in terms of the 63 NG bosons. 
It is further straightforwardly extended to a scale-invariant version so as to incorporate the TD, $\phi$, 
as a composite Higgs~\cite{Matsuzaki:2012vc}. 
The chiral perturbation theory (ChPT) of the scale-invariant version can also be formulated by assigning 
the chiral order counting $m_\phi^2 ={\cal O} (p^2)$~\cite{Matsuzaki:2013eva}.

Three of the 63 NG bosons are eaten by the SM weak gauge bosons, 
when the SM gauge interactions are switched on, while the other 60 remain as physical states (``technipions"), 
all acquiring mass to become pseudo-NG bosons by the SM gauging and the ETC gauging, 
which explicitly breaks the chiral symmetry $G$ down to the symmetry corresponding to the EW symmetry. 
The gauge couplings of these explicit breakings are small and perturbative, and therefore the masses of the technipions may be estimated by the Dashen formula 
at the lowest-order perturbation, just like the estimate of the $\pi^+ - \pi^0$ mass difference in QCD. 
It turns out that  the masses of all the technipions are drastically enhanced by the walking dynamics of WTC~\cite{Yamawaki:1985zg,Holdom:1987ed, Harada:2005ru}. 
In the case of the one-family model, the masses of the walking technipions were explicitly estimated~\cite{Jia:2012kd, Kurachi:2014xla} 
to be of ${\cal O}$ (TeV) (see Sec.~\ref{sec:properties} below),
suggesting a new possibility that the technirhos decay directly to the SM particles, rather than through 
the technipions.

In this paper, we consider another type of the technihadrons -- 
vector resonances (here we are confined to the flavor-nonsinglet ones, called ``technirho mesons") -- 
which are expected to exist as typical bound states in the generic dynamical EW symmetry-breaking scenarios, 
not restricted to WTC or the one-family model. 
In the case of WTC we extend  the scale-invariant version of the low-energy effective theory~\cite{Matsuzaki:2012gd,Matsuzaki:2012vc,Matsuzaki:2012mk} -- i.e., 
the case of the one-family model based on $G/H= SU(8)_L \times SU(8)_R/SU(8)_V$~\cite{Jia:2012kd} -- 
 in a way that includes the technirhos by using the hidden local symmetry (HLS)~\cite{Bando:1984ej,
Harada:2003jx}~\footnote{
The flavor-singlet technivector meson can be incorporated into the HLS Lagrangian by taking  $G/H= U(8)_L \times U(8)_R/U(8)_V$ 
instead of  $G/H= SU(8)_L \times SU(8)_R/SU(8)_V$. 
As we discuss later, however, it would be less interesting compared with the technirho in the LHC phenomenology and 
is not discussed in this paper. }. 
Similar discussions on the technirho's based on the HLS were  done for the one-doublet model with $G/H=SU(2)_L \times SU(2)_R/SU(2)_V$ 
without the scale symmetry/TD [the ``breaking electroweak symmetry strongly" (BESS) model]~\cite{Casalbuoni:1985kq}. 
This is the first study of the HLS for the one-family model 
as well as its scale-invariant version, which implies a novel salient LHC phenomenology 
involving the TD and the colored technirho.  
It is to be noted that the ChPT was formulated for the HLS Lagrangian~\cite{Harada:2003jx} 
and can be extended to the scale-invariant version of the HLS in the same way as in the case without the HLS~\cite{Matsuzaki:2013eva}, 
although we do not include the loop effects in this paper. 
Based on the scale-invariant HLS, we first evaluate the constraint on the technirho meson masses from the current LHC data, and then future LHC phenomenology 
is discussed. 
We find that the color-octet technirho produced via the Drell-Yan (DY) process which decays to the TD along with the gluon 
is especially interesting as a discovery channel at the LHC to probe the one-family WTC.

The paper is organized as follows.
In Sec.~II, we formulate the scale-invariant HLS 
based on the manifold 
$[SU(8)_L \times SU(8)_R]_{\rm global} \times SU(8)_{\rm local}/SU(8)_V$, including the TD, technipions, and 
technirho mesons. 
In Sec.~III the decay widths and branching ratios of the technirho mesons in the one-family WTC are discussed. 
In Sec.~IV we explore the LHC phenomenology of the technirho mesons and place limits on the masses 
from the currently available LHC data on searches for new spin-1 resonances. 
We then discuss the discovery channels of the technirho mesons, which include the TD 
as the daughter particle of the parent technirho mesons.  
A Summary is given in Sec.~V.  
The explicit forms of the technirho couplings relevant to the LHC phenomenology 
and the partial decay widths are presented in Appendices A and B, respectively.

\section{The scale-invariant HLS for walking technirho mesons  
of the one-family model } 

In this section, based on the HLS formalism~\cite{Bando:1984ej,Harada:2003jx},  we give a formulation of the scale-invariant HLS 
for the one-family WTC model, which includes the TD and technipions, as well as the technirho mesons as the HLS gauge bosons. 
The HLS formalism makes it straightforward to simultaneously incorporate both the external (SM) gauge and HLS (vector-meson) interactions, 
in contrast to other approaches for the inclusion of the vector mesons into the chiral Lagrangian. 
It actually turns out to be crucial for studying the LHC phenomenology, as will be seen below. 
Furthermore, the ChPT for the systematic loop expansion has been developed only in the HLS formalism, although the vector mesons can also be incorporated into the chiral Lagrangian by other formalisms which are equivalent to the HLS Lagrangian at the on-shell tree level~\cite{Harada:2003jx}.

The Lagrangian for the technipion is expressed as the usual nonlinear sigma model based on the manifold $G/H = SU(8)_L \times SU(8)_R/SU(8)_V$. 
The TD is incorporated by forcing the chiral effective theory to be scale-invariant 
through the introduction of the compensating nonlinear field $\chi(x)=e^{\phi(x)/F_\phi}$, 
which transforms under scale transformation as 
$\delta \chi(x) = (1 + x^\nu \partial_\nu) \chi(x)$,  
so that $\phi(x)$ does scale nonlinearly as $\delta \phi(x)= F_{\phi} + x^\nu \partial_\nu \phi(x)$,
where $\phi(x)$ and $F_\phi$ are the TD field and its decay constant, respectively~\cite{Matsuzaki:2012gd,Matsuzaki:2012vc,Matsuzaki:2012mk}. 
The resultant one-family scale-invariant action is explicitly given by the Lagrangian~\cite{Jia:2012kd}  
\begin{eqnarray} 
{\cal L}
= \frac{F_{\pi}^2}{4} \cdot \chi^2 (x)\cdot {\rm tr}[{\cal D}_\mu U^\dag {\cal D}^\mu U] + \frac{F_\phi^2}{2}\partial_\mu\chi(x)\partial^\mu\chi(x)\,,
\label{Lag:int} 
\end{eqnarray}
where $U(x)=e^{ i \frac{2\pi(x)}{F_\pi}} $ (with $F_\pi$ being the decay constant of the NG  bosons) transforms as $U\rightarrow g_L \cdot U \cdot g_R^\dagger$, with $(g_L, g_R) \in G=SU(8)_L\times SU(8)_R$, as 
does the covariant derivative ${\cal D}^\mu U(x) =\partial^\mu U(x) -i {\cal L}^\mu(x) U(x) +i U(x) {\cal R}^\mu(x)$, where
 $G$ is formally fully gauged by the external gauge fields $({\cal L}^\mu(x), {\cal R}^\mu(x))$. The action for the Lagrangian (\ref{Lag:int}) is invariant under the gauged-$G$ symmetry as well as the scale symmetry. In the realistic application to WTC, the external gauge fields are restricted to the SM gauge fields of $SU(3) \times SU(2)_L \times U(1)_Y$.  Besides the scale-invariant term, there exists a scale-anomaly term which reproduces the TD mass and as well as terms involving the TD coupling 
to the SM fields~\cite{Matsuzaki:2012vc}.

Now, it is straightforward to introduce the technirho into the Lagrangian (\ref{Lag:int}) in the standard manner of the HLS formalism~\cite{Bando:1984ej}.  The HLS can be made explicit by writing $U(x)=e^{ i \frac{2\pi(x)}{F_\pi}} =\xi_L^\dagger(x) \cdot \xi_R(x)$, 
where the $\xi_{L,R}$ are parametrized as 
\begin{equation} 
 \xi_{L,R}(x) = e^{\frac{i \sigma(x)}{F_\sigma}} e^{\mp  \frac{i \pi(x)}{F_\pi}}  
 \,, 
 \qquad 
 ({\pi(x)=\pi^A(x) X^A}\,, \qquad {\sigma(x) = \sigma^A(x) X^A} )
\,, 
\end{equation} 
with the broken generators $X^A$ and  
the fictitious NG bosons $\sigma^A(x)$ (not to be confused with the scalar meson) along with the decay constant $F_\sigma$, which are 
to be absorbed into the HLS. 
The $ \xi_{L,R}(x)$ transform as $\xi_{L,R} \rightarrow h(x) \cdot \xi_{L,R} \cdot g^\dagger_{L,R}$ under $ G_{\rm global}\times H_{\rm local} =[SU(8)_L\times SU(8)_R]_{\rm global} \times SU(8)_{\rm local}$, where $h\in H_{\rm local} =SU(8)_{\rm local}$ is the HLS, and $g_{L,R} \in G_{\rm global}=[SU(8)_L\times SU(8)_R]_{\rm global}$.
When we fix the gauge [unitary gauge $\sigma(x)=0$] as $\xi_L^\dagger(x) =\xi_R(x)=\xi(x)=e^{ i \frac{\pi(x)}{F_\pi}} $,  $H_{\rm local}$ and $H_{\rm global}(\subset G_{\rm global})$ are both spontaneously broken down to a single $H$ which is a diagonal sum of both of them, and accordingly $G_{\rm global}$ is reduced back to the original chiral symmetry $G$ in the model based on $G/H$: $\xi$ transforms as $\xi \rightarrow h(g, \pi)\, \xi\, g^\dagger_{L,R}$, with $h(g, \pi)$ being the $\pi(x)$-dependent (global) $H$ transformation of $G/H$.

The technirho mesons are introduced as the gauge bosons 
of the HLS 
$H_{\rm local} =
SU(8)_{\rm local}$ through the covariant derivative 
 $D_\mu \xi_{L,R} (x)= \partial_\mu \xi_{L,R}(x) - i V_\mu(x) \xi_{L,R}(x) + i \xi_{L,R} {\cal L}_\mu(x)({\cal R}_\mu(x))
$, with the HLS gauge field $V_\mu$  and the external gauge fields ${\cal L}_\mu$ and ${\cal R}_\mu$. 
$G_{\rm global}$ is again fully gauged for formal discussion to make the invariance transparent. 
The resulting form of the Lagrangian is as follows:
\begin{equation} 
 {\cal L} =  \chi^2(x) \cdot \left( F_\pi^2 {\rm tr}[ \hat{\alpha}_{\mu \perp}^2] + F_\sigma^2 {\rm tr}[\hat{\alpha}_{\mu ||}^2]  \right) 
  - \frac{1}{2g^2} {\rm tr}[V_{\mu\nu}^2] 
  \,, \label{Lag}
\end{equation}
where
\begin{equation}
  \hat{\alpha}_{\mu \perp,||} = \frac{ D_\mu \xi_R \cdot \xi_R^\dag \mp D_\mu \xi_L \cdot \xi_L^\dag}{2i} 
\,. 
\end{equation}
The covariantized Maurer-Cartan 1-forms $\hat{\alpha}_{\mu \perp,||}$ transform  as
$\hat{\alpha}_{\mu \perp,||} \rightarrow h(x)\cdot \hat{\alpha}_{\mu \perp,||}\cdot h^\dag(x) $. 
 Without the kinetic term of the HLS gauge fields $V_\mu(x)$ (namely by integrating out the $V_\mu$), 
the Lagrangian is reduced 
to the nonlinear sigma model based on $G/H$ in the unitary gauge $\sigma(x)=0$ ($\xi_L^\dagger(x) =\xi_R(x)=\xi(x)=e^{ i \frac{\pi(x)}{F_\pi}} $). 

The 63 chiral NG bosons are embedded in the adjoint representation of the $SU(8)$ group~\cite{Jia:2012kd}: 
\begin{eqnarray} 
\sum_{A=1}^{63} \pi^A(x) X^A 
&=& 
\sum_{i=1}^3 \Pi^i(x) X_{\Pi}^i  
+ 
\sum_{i=1}^3 P^i(x) X^i_{P} + P^0 (x) X_{P} 
\nonumber \\  
&& 
+ \sum_{i=1}^3 \sum_{a=1}^8 \theta^i_a (x) X_{\theta a}^i 
+ 
\sum_{a=1}^8 \theta_a^0 (x) X_{\theta a}  
\nonumber \\ 
&& 
+ 
\sum_{c=r,g,b} \sum_{i=1}^3 \left[ T_c^{i}(x) X_{T c}^{i} + \bar{T}_c^{i}(x) X_{\bar{T} c}^{i} \right] 
+
\sum_{c=r,g,b} \left[ T_c^0 (x) X_{T c}  + \bar{T}_c^0 (x) X_{\bar{T} c} \right] 
\,, 
\end{eqnarray} 
where ($\tau^i=\sigma^i/2$)
\begin{eqnarray} 
 X_{\Pi}^i 
 &=& 
\frac{1}{2} \left( 
\begin{array}{c|c} 
  \tau^i \otimes {\bf 1}_{3\times 3}  &  \\ 
  \hline 
   & \tau^i 
\end{array}
\right)  \,, 
\qquad 
X^i_P 
= 
\frac{1}{2 \sqrt{3}}
\left( 
\begin{array}{c|c} 
  \tau^i \otimes {\bf 1}_{3\times 3}  &  \\ 
  \hline 
   & -3 \cdot \tau^i 
\end{array}
\right)  \,, \qquad 
X_P
= 
\frac{1}{4 \sqrt{3}}
\left( 
\begin{array}{c|c} 
  {\bf 1}_{6 \times 6}  &  \\ 
  \hline 
   & -3\cdot {\bf 1}_{2 \times 2} 
\end{array}
\right) 
\,, \nonumber \\ 
 X_{\theta a}^i 
 &=& 
\frac{1}{\sqrt{2}}  \left( 
\begin{array}{c|c} 
  \tau^i \otimes \lambda_a  &  \\ 
  \hline 
   & 0 
\end{array}
\right)  \,, 
\qquad 
X_{\theta_a} 
= 
\frac{1}{2 \sqrt{2}}
\left( 
\begin{array}{c|c} 
  {\bf 1}_{2\times 2} \otimes \lambda_a  &  \\ 
  \hline 
   & 0  
\end{array}
\right)  \,, \nonumber \\ 
X^{i}_{T c} 
&=&  
\frac{1-i}{2}
\left( 
\begin{array}{c|c} 
 & \tau^i \otimes {\bf e}_c    \\ 
  \hline 
\tau^i \otimes {\bf e}_c^\dag  &  
\end{array}
\right) 
\,, \qquad 
X^{i}_{\bar{T} c}  
= 
\left( X^{i}_{T c}  \right)^\dagger
\,, \nonumber \\ 
X_{T c}  
&=&  
\frac{1-i}{4}
\left( 
\begin{array}{c|c} 
 & {\bf 1}_{2 \times 2} \otimes {\bf e}_c    \\ 
  \hline 
{\bf 1}_{2 \times 2} \otimes {\bf e}_c^\dag  &  
\end{array}
\right) 
\,, \qquad 
X_{\bar{T} c}  
= 
\left( X_{T c}  \right)^\dagger
\,, 
\end{eqnarray}
with 
${\bf e}_c$ being a three-dimensional unit vector in color space  
and the generators are normalized as ${\rm Tr}[X^A X^B]=\delta^{AB}/2$. 
Among the above, $\Pi^i$ become longitudinal degrees of freedom of the SM $W^{\pm}$ and $Z$ bosons. 
It is convenient to express $\pi$ in a blocked $8 \times 8$ matrix form as 
\begin{equation} 
 \pi^A X^A = 
 \Bigg(
 \begin{array}{c|c} 
(\pi_{QQ})_{6 \times 6} & (\pi_{QL})_{2 \times 6} \\ 
\hline  
 (\pi_{LQ})_{6 \times 2} & (\pi_{LL})_{2 \times 2} 
 \end{array} 
 \Bigg) 
 \,, \label{pi:para}
\end{equation}
where 
\begin{eqnarray} 
\pi_{QQ} &=& \left[ \sqrt{2} \theta  + \frac{1}{\sqrt{2}} \theta^0   \right] 
+ \left(  \frac{1}{2} \Pi  + \frac{1}{2\sqrt{3}} P  + \frac{1}{4 \sqrt{3}} P^0 \right) \otimes {\bf 1}_{3 \times 3}
\,, \nonumber \\ 
\pi_{QL} &=&   T + \frac{1}{2} T^0  
\,, \nonumber \\ 
\pi_{LQ} &=& \pi_{QL}^\dag = 
\bar{T}  + \frac{1}{2} \bar{T}^0  
\,, \nonumber \\ 
\pi_{LL} &=& 
 \left( \frac{1}{2} \Pi  - \frac{\sqrt{3}}{2} P  - \frac{\sqrt{3}}{4} P^0   \right) 
 \,, \nonumber \\
\theta &=& \theta_a^i \tau^i \frac{\lambda_a}{2} 
\,, \qquad 
\theta^0 = \theta_a^0  \cdot {\bf 1}_{2 \times 2} \cdot \frac{\lambda_a}{2} 
\,, \nonumber \\ 
T &=& T_c^i {\bf e}_c \tau^i 
\,, \qquad 
T^0 = T_c^0 {\bf e}_c 
\,, \nonumber \\ 
P &=&  P^i \tau^i 
\,, \qquad 
P^0 = P^0 \cdot {\bf 1}_{2 \times 2} 
\,, \nonumber \\ 
\Pi &=& \Pi^i \tau^i 
\,.   \nonumber 
\end{eqnarray}

The technirho meson fields are also parametrized in a way similar to $\pi$: 
\begin{eqnarray} 
\sum_{A=1}^{63} \rho^A(x) X^A 
&=& 
\sum_{i=1}^3 \rho_{\Pi}^i  (x) X_{\Pi}^i  
+ 
\sum_{i=1}^3  \rho_{P}^i (x) X^i_{P} +  \rho_{P}^0  (x) X_{P} 
\nonumber \\  
&& 
+ \sum_{i=1}^3 \sum_{a=1}^8  \rho_{\theta a}^i  (x) X_{\theta a}^i 
+ 
\sum_{a=1}^8 \rho_{\theta_a}^0 (x) X_{\theta a}  
\nonumber \\ 
&& 
+ 
\sum_{c=r,g,b} \sum_{i=1}^3 \left[ \rho_{T c}^{i}(x) X_{T c}^{i} + \bar{\rho}_{T c}^{i}(x) X_{\bar{T} c}^{i} \right] 
+
\sum_{c=r,g,b} \left[ \rho_{T c}^0(x) X_{T c}  + \bar{\rho}_{T_c}^0(x) X_{\bar{T} c} \right] 
\,. \label{rho-parametrize}
\end{eqnarray}  
They are embedded in a  $8 \times 8$ block-diagonal form, $V_\mu=V^A_\mu X^A$, as 
\begin{equation} 
 \rho^\mu  = \frac{V^{\mu A} X^A}{g} = 
 \Bigg(
 \begin{array}{c|c} 
(\rho^\mu_{QQ})_{6 \times 6} & (\rho^\mu_{QL})_{2 \times 6} \\ 
\hline  
 (\rho_{LQ}^\mu)_{6 \times 2} & (\rho^\mu_{LL})_{2 \times 2} 
 \end{array} 
 \Bigg) 
 \,, \label{rho:para}
\end{equation}
with 
\begin{eqnarray} 
\rho^\mu_{QQ} &=& \left[ \sqrt{2} \rho^{\mu}_\theta + \frac{1}{\sqrt{2}} \rho^{\mu 0}_\theta   \right]  
+ \left(  \frac{1}{2} \rho^{\mu}_{\Pi}  + \frac{1}{2\sqrt{3}} \rho^{\mu}_P  + \frac{1}{4 \sqrt{3}} \rho^{0 \mu}_P  \right) \otimes {\bf 1}_{3\times 3} 
\,, \nonumber \\ 
\rho_{QL}^\mu &=& \rho^\mu_T + \frac{1}{2} \rho^{0\mu}_T
\,, \nonumber \\ 
\rho^\mu_{LQ} &=& (\rho^\mu_{QL})^\dag = 
\bar{\rho}^\mu_T  + \frac{1}{2} \bar{\rho}^{0\mu}_T  
\,, \nonumber \\ 
\rho^\mu_{LL} &=& 
 \left( \frac{1}{2} \rho^\mu_\Pi  - \frac{\sqrt{3}}{2} \rho^\mu_P - \frac{\sqrt{3}}{4} \rho^{0\mu}_P \right) 
 \,,\nonumber \\
 \rho_{\theta}^\mu &=& \rho_{\theta a}^{i\mu} \tau^i \frac{\lambda_a}{2} 
\,, \qquad 
\rho^{0\mu}_\theta = \rho_{\theta a}^{0\mu}  \cdot {\bf 1}_{2 \times 2} \cdot \frac{\lambda_a}{2} 
\,, \nonumber \\ 
\rho^\mu_T &=& \rho_{Tc}^{i\mu} {\bf e}_c \tau^i 
\,, \qquad 
\rho^{0\mu}_T = \rho_{Tc}^{0\mu} {\bf e}_c 
\,, \nonumber \\ 
\rho_P^\mu &=&  \rho_P^{i\mu} \tau^i 
\,, \qquad 
\rho_P^{0\mu} = \rho_P^{0\mu} \cdot {\bf 1}_{2 \times 2} 
\,, \nonumber \\ 
\rho^\mu_\Pi &=& \rho^{i\mu}_\Pi \tau^i 
\,. \nonumber 
\end{eqnarray}
Here we used the same basis of the $SU(8)_V$ matrix as that of $\pi$ since 
$\rho_\Pi^i$ are produced only by the Drell-Yan processes through the mixing with $W$ and $Z$ bosons, 
which absorb $\Pi^i$. With this base, the other color-singlet isotriplet technirhos, $\rho_P^i$, 
are not produced due to the orthogonality of ${\rm tr}[X_{\Pi}^i X_P^j]=0$, as will be seen later.

The external gauge fields ${\cal L}_\mu$ and ${\cal R}_\mu$ involve the $SU(3)_c$, $SU(2)_W$ and $U(1)_Y$ gauge fields 
$(G_\mu, W_\mu, B_\mu)$ in the SM as follows: 
\begin{eqnarray} 
{\cal L}_\mu 
&=& 
2 g_W W_\mu^i X^i_{\Pi} + \frac{2}{\sqrt{3}} g_Y B_\mu X_P  + \sqrt{2} g_s  G_\mu^a X_{\theta_a}  
\,, \nonumber \\ 
{\cal R}_\mu 
&=&
2 g_Y B_\mu \left( X_{\Pi}^3 + \frac{1}{\sqrt{3}} X_P \right) + \sqrt{2} g_s G_\mu^a X_{\theta_a} 
\,. 
\end{eqnarray}
Through the standard diagonalization procedure, the left and right gauge fields are 
expressed in terms of the mass eigenstates $(W^\pm, Z, \gamma, g)$ as 
\begin{eqnarray} 
{\cal L}_\mu 
&=& 
g_s G_\mu^a \Lambda^a + e Q_{\rm em} A_\mu + \frac{e}{sc} \left( I_3 - s^2 Q_{\rm em} \right) Z_\mu + 
\frac{e}{\sqrt{2}s} \left( W_\mu^+ I^+ + {\rm H.c.} \right)
\,, \nonumber \\ 
{\cal R}_\mu 
&=&
g_s G_\mu^a \Lambda^a + e Q_{\rm em} A_\mu  - \frac{es}{c} Q_{\rm em} Z_\mu  
\,, 
\end{eqnarray}
where $s$ ($c^2=1-s^2$) denotes the standard weak mixing angle defined by $g_W=e/s$ and $g_Y=e/c$, and  
\begin{eqnarray} 
\Lambda_a &=& \sqrt{2} \, X_{\theta_a} \,, \qquad  
I_3 = 2 \, X_{\Pi}^3 \,  , \qquad 
Q_{\rm em} = I_3 + Y \,, \nonumber \\ 
Y&=& \frac{2}{\sqrt{3}} X_P  \,, \qquad 
I_+ = 2(X_{\Pi}^1 + i X_{\Pi}^2)   \,, \qquad 
I_- = (I_+)^\dag 
\,. 
\end{eqnarray}
It is convenient to define the vector and axial-vector gauge fields ${\cal V}_\mu$ and ${\cal A}_\mu$ as
\begin{equation} 
 {\cal V}_\mu = \frac{{\cal R}_\mu + {\cal L}_\mu }{2} 
 \,,\qquad
 {\cal A}_\mu = \frac{{\cal R}_\mu - {\cal L}_\mu}{2} 
\,, 
\end{equation}  
so that they are expressed in a blocked-$8 \times 8$ matrix form:  
\begin{eqnarray} 
 {\cal V}^\mu &=& 
 \Bigg( 
 \begin{array}{c|c} 
 ({\cal V}^\mu_{QQ})_{6 \times 6} & {\bf 0}_{2 \times 6} \\ 
 {\bf 0}_{6 \times 2} & ({\cal V}^\mu_{LL})_{2 \times 2}
\end{array} 
 \Bigg) 
 \,,\label{v:para} \\ 
  {\cal A}^\mu &=& 
 \Bigg( 
 \begin{array}{c|c} 
 ({\cal A}^\mu_{QQ})_{6 \times 6} & {\bf 0}_{2 \times 6} \\ 
 {\bf 0}_{6 \times 2} & ({\cal A}^\mu_{LL})_{2 \times 2}
\end{array} 
 \Bigg) 
 \,, \label{v:a:para}
\end{eqnarray}
where 
\begin{eqnarray} 
 {\cal V}^\mu_{QQ} &=& 
{\bf 1}_{2 \times 2} \cdot g_s G^\mu_a \frac{\lambda_a}{2}   
+ 
\left[ 
 e Q_{\rm em}^q A^\mu + \frac{e}{2sc} z_V^q Z^\mu + \frac{e}{2\sqrt{2}s} \left( \tau^+ W^{\mu +} + {\rm H.c.} \right)
\right] \cdot {\bf 1}_{3 \times 3} 
\,, \nonumber \\ 
 {\cal V}^\mu_{LL} &=& 
\left[ 
 e Q_{\rm em}^l A^\mu + \frac{e}{2sc} z_V^l Z^\mu + \frac{e}{2\sqrt{2}s} \left( \tau^+ W^{\mu +} + {\rm H.c.} \right)
\right]
\,, \nonumber \\ 
{\cal A}^\mu_{QQ} &=& 
- \left(
 \frac{e}{2sc} \tau^3 Z^\mu + \frac{e}{2 \sqrt{2}s}\left( \tau^+ W^{\mu +} + {\rm H.c.} \right) 
\right) \cdot {\bf 1}_{3 \times 3} 
\,, \nonumber \\
{\cal A}^\mu_{LL} &=& 
- \left(
 \frac{e}{2sc} \tau^3 Z^\mu + \frac{e}{2 \sqrt{2}s}\left( \tau^+ W^{\mu +} + {\rm H.c.} \right) 
\right) 
\,, \nonumber \\
Q_{\rm em}^{q} &=& 
\left(
\begin{array}{cc}
 2/3 & 0 \\ 
 0 & -1/3 
\end{array}
\right) 
\,, \qquad 
Q_{\rm em}^{l} = 
\left(
\begin{array}{cc}
 0 & 0 \\ 
 0 & -1 
\end{array}
\right) 
\,, \nonumber \\ 
z_V^{q,l} &=& \tau^3 - 2 s^2 Q_{\rm em}^{q,l} 
\,, \qquad 
\tau^+ = 
\left( 
 \begin{array}{cc}
 0 & 1 \\ 
 0 & 0 
\end{array}
\right) 
\,, \qquad 
\tau^- = (\tau^+)^\dag  
\,. \nonumber 
\end{eqnarray}

Interactions among technihadrons and SM particles can be obtained by expanding the Lagrangian in Eq.(\ref{Lag}) in powers of $\pi$ and $\phi$ with 
the HLS fixed as unitary gauge $(\sigma=0)$. 
Basic HLS relations include~\cite{Bando:1984ej}
\begin{eqnarray}
M_\rho^2 &=& a g^2 F_\pi^2, \label{eq:Mrho} \\
g_{\rho\pi\pi} &=& \frac{a}{2}\, g, \\
g_{{\cal V} \pi\pi} &=& \left(1-\frac{a}{2}\right),
\label{eq:Vpipi}
\end{eqnarray}
with 
\begin{equation}
a \equiv \frac{F_\sigma^2}{F_\pi^2},
\end{equation} 
where $M_\rho$, $g_{\rho\pi\pi}$ and $g_{\gamma \pi\pi}$ have been read off from the following terms: 
\begin{eqnarray}
{\cal L}_{M_\rho^2} &=& M_\rho^2\, {\rm tr}\left[ \rho_\mu \rho^\mu \right], \\
{\cal L}_{{\cal V}\pi\pi} &=& 2\, i\, g_{{\cal V}\pi\pi}\, {\rm tr} \left[ {\cal V}^\mu [\partial_\mu \pi, \pi] \right], \\
{\cal L}_{\rho\pi\pi} &=& 2\, i\, g_{\rho\pi\pi}\, {\rm tr} \left[ \rho^\mu [\partial_\mu \pi, \pi] \right].
\end{eqnarray}
Note that, from Eq.~(\ref{eq:Vpipi}), direct couplings of SM gauge bosons to two pions vanish when we take $a=2$:
\begin{equation} 
  g_{{\cal V}\pi\pi} = 0 \, \qquad  {\rm for} \qquad  a=2
  \,. \label{VMD:def}
\end{equation}
In that case, the couplings of the SM gauge bosons (${\cal V}_\mu$) to two pions arise only from the $\rho$-${\cal V}$ mixing 
(vector-meson dominance).

The explicit forms of interactions relevant to the current study are summarized in Appendix~\ref{app:int}.
Among these interaction terms, the most relevant terms in the present study 
are ${\cal V}$-$\rho$ and 
${\cal V}$-$\rho$-$\phi$ vertices. 
These terms arise from the $\chi^2 F_\sigma^2 {\rm tr}[\hat{\alpha}_{\mu ||}^2]$ term in Eq.(\ref{Lag}) 
by expanding $\chi$ and $\hat{\alpha}_{\mu ||}$ in terms of the technidilaton ($\phi$) and pion $(\pi)$ fields:  
\begin{eqnarray} 
\chi^2 F_\sigma^2 {\rm tr}[\hat{\alpha}_{\mu||}^2]
&=& F_\sigma^2 \left( 1 + \frac{2\phi}{F_\phi} + \cdots \right) 
\times \nonumber \\ 
&& 
{\rm tr} \left[ 
 ({\cal V}_\mu - V_\mu)^2 + \frac{i}{F_\pi^2} V_\mu [\partial^\mu\pi, \pi] + \frac{2i}{F_\pi} V_\mu [{\cal A}^\mu, \pi] 
 + \cdots
\right]
\,. 
\end{eqnarray} 
 From the first term in the square bracket, and by using Eqs.(\ref{rho:para}) and (\ref{v:para}), one can readily read off the ${\cal V}$-$\rho$ 
and ${\cal V}$-$\rho$-$\phi$  vertices as 
\begin{eqnarray} 
{\cal L}_{{\cal V} \rho}  
&=& 
 - 2 g F_\sigma^2 {\rm tr}[  {\cal V}_\mu \rho^\mu ] 
\nonumber\\ 
&=&
 - 2 g F_\sigma^2
 \Bigg[ 
  \frac{g_s}{\sqrt{2}}  G_\mu^a \rho_{\theta}^{0 a \mu}
  + 
  e A_\mu \left\{
 \rho^{3 \mu}_{\Pi} + \frac{1}{\sqrt{3}} \rho^{0 \mu}_{P} 
 \right\}  
 \nonumber \\ 
&&
+ \frac{e}{2sc} Z_\mu  \left\{ 
  (c^2-s^2) \rho^{3 \mu}_\Pi - \frac{2}{\sqrt{3}} s^2 \rho^{0 \mu}_{P} 
 \right\} 
 + \frac{e}{2 s} 
 \left\{ 
  W_\mu^- \rho^{\mu+}_{\Pi} + {\rm H.c.} 
 \right\}
 \Bigg]
 \,, \label{v-rho:mixing:0}
\end{eqnarray}
and 
\begin{eqnarray} 
{\cal L}_{{\cal V} \rho \phi}  
&=& 
 - \frac{4 g F_\sigma^2}{F_\phi} \phi {\rm tr}[  {\cal V}_\mu \rho^\mu ] 
\nonumber\\ 
&=&
 - \frac{4 g F_\sigma^2}{F_\phi} \phi 
 \Bigg[ 
  \frac{g_s}{\sqrt{2}}  G_\mu^a \rho_{\theta}^{0 a \mu}
  + 
  e A_\mu \left\{
 \rho^{3 \mu}_{\Pi} + \frac{1}{\sqrt{3}} \rho^{0 \mu}_{P} 
 \right\}  
 \nonumber \\ 
&&
+ \frac{e}{2sc} Z_\mu  \left\{ 
  (c^2-s^2) \rho^{3 \mu}_\Pi - \frac{2}{\sqrt{3}} s^2 \rho^{0 \mu}_{P} 
 \right\} 
 + \frac{e}{2 s} 
 \left\{ 
  W_\mu^- \rho^{\mu+}_{\Pi} + {\rm H.c.} 
 \right\}
 \Bigg]
 \,. \label{TD-rho-V}
\end{eqnarray}
Here we have defined the charged rho-meson fields as 
\begin{equation} 
 \rho^{\mu \pm}_{\Pi} = \frac{\rho^{1 \mu}_{\Pi} \mp i \rho^{2 \mu}_{\Pi} }{\sqrt{2}}
\,. 
\end{equation} 
 Note the absence of $A-\rho_P^3, Z-\rho_P^3$ and $W^\pm-\rho_P^\mp$, 
$A-\rho_P^3-\phi, Z-\rho_P^3-\phi$, and $W^\pm-\rho_P^\mp-\phi$ terms due to 
the orthogonality of the $SU(8)_V$ generators.  
As will be discussed more explicitly, the terms in Eq.~(\ref{v-rho:mixing:0})  
are crucial for technirho-meson productions through the Drell-Yan process and 
decays to the SM fermions via the vector-meson dominance, as in Eq.(\ref{VMD:def}). 
The terms in Eq.~(\ref{TD-rho-V}) are relevant for decays which involve 
the Higgs boson (TD).

\section{Decay widths and branching ratios of technirho mesons} 
\label{sec:properties}

\begin{figure}[t]
\begin{center}
   \includegraphics[scale=0.6]{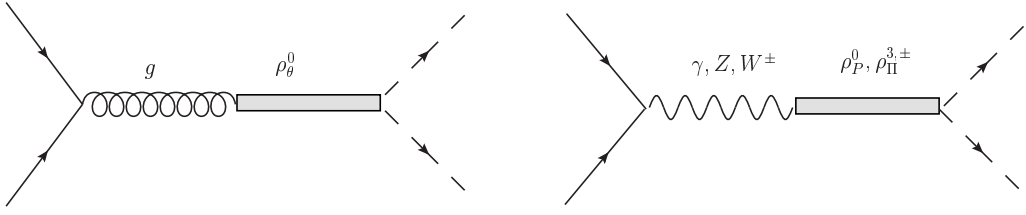}
\caption{ An illustration of the relevant LHC production and two-body decay processes for $\rho_{\theta}^0, \rho_P^0, \rho^{\pm ,3}_\Pi$.   
\label{fig:DY}
}
\end{center} 
\end{figure} 

Having formulated the low-energy effective Lagrangian of the one-family model and derived relevant interactions among technihadrons and SM fields, in this section we study the decay widths and branching ratios of the technirho mesons and related collider phenomenology that are based on them. 
The Lagrangian in Eq.(\ref{Lag}) has four parameters: $F_\pi$, $F_\sigma$, $F_\phi$, and $g$. 
We fix $F_\pi = 123$ GeV, which is set by the EW scale $v_{\rm EW} \simeq 246$ GeV through the relation 
\begin{equation}
F_\pi= \frac{v_{\rm EW}}{\sqrt{N_{D}}} \simeq 123 \ {\rm GeV\ \ (for\ } N_D=4).  
\end{equation}
 As for the TD decay constant $F_\phi$, 
we use the best-fit value by which the TD can be fitted well to the current LHC Higgs data (see Ref.~\cite{Matsuzaki:2012mk}): 
\begin{equation} 
F_\phi=F_{\phi}|_{\rm best} \simeq v_{\rm EW}/0.22 
\,, 
\end{equation} 
Also, we impose the vector-meson dominance, which is achieved by taking $a= F_\sigma^2/F_\pi^2$ to be 
[see Eq.~(\ref{VMD:def})] 
\begin{equation} 
 a = 2 
 \,. 
\end{equation}
The remaining HLS parameter, $g$, will be fixed by the input value of the technirho mass, $M_\rho$, through the relation in Eq.~(\ref{eq:Mrho}).

In the next section, we will study technirho production through the mixing with the SM gauge bosons which are produced by the Drell-Yan process~\footnote{It should be noted that we do not include the gluon-gluon fusion process for $\rho_{\theta}^0$ production since there is an accidental cancellation of the 
$g-g-\rho_{\theta}^0$ on-shell amplitude by the presence of contribution from the non-Abelian 
$\rho_{\theta}^0-\rho_{\theta}^0-\rho_{\theta}^0$ vertex~\cite{Zerwekh:2001uq,Chivukula:2001gv}. (This is true for the leading and the next-to-leading order in the derivative expansion of the Lagrangian in Eq.~(\ref{Lag}).) Also, we have no $g-\rho_{T}-\rho_{T}$ vertex at the leading order due to 
the $SU(8)_V$ invariance, so the current limit~\cite{Abazov:2009ab} on vector leptoquarks through the Drell-Yan process is not 
applicable to $\rho_{T}$. }.  
The types of technirho mesons produced by such a process are $\rho_{\theta}^0$, $\rho_P^0$ and $\rho^{\pm ,3}_\Pi$, as illustrated in Fig.~\ref{fig:DY}. 
Note that $\rho_P^{\pm, 3}$ are not produced via the Drell-Yan process because 
they do not mix with the SM gauge bosons due to the orthogonality of the $SU(8)_V$ symmetry [see Eq.(\ref{v-rho:mixing}) in Appendix~A].    
Thus, in this section, we show several coupling properties of $\rho_{\theta}^0$, $\rho_P^0$, and $\rho^{\pm ,3}_\Pi$.

\begin{figure}[ht]
\begin{center}
   \includegraphics[width=8.5cm]{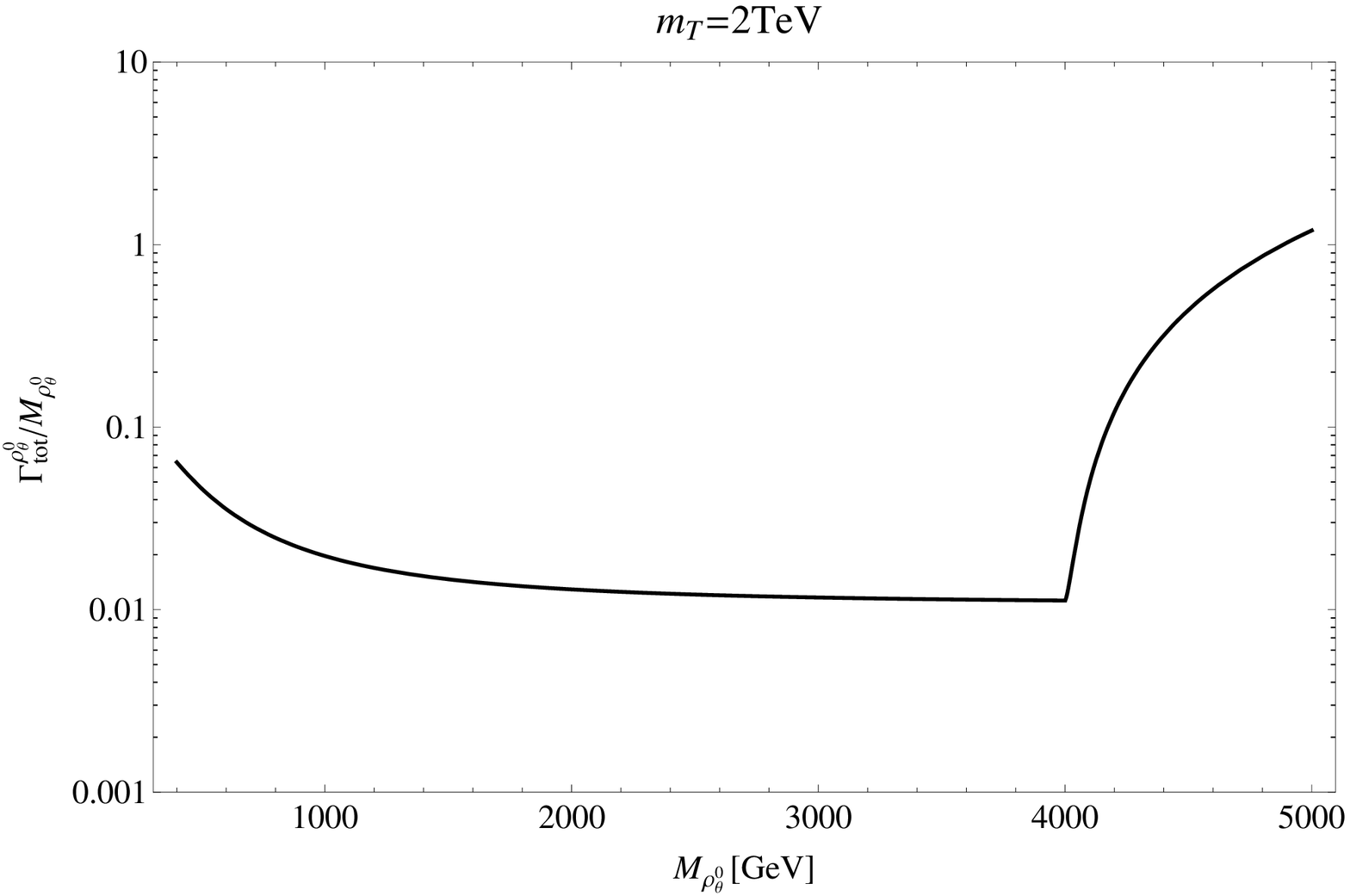}\ \ \ 
   \includegraphics[width=8.5cm]{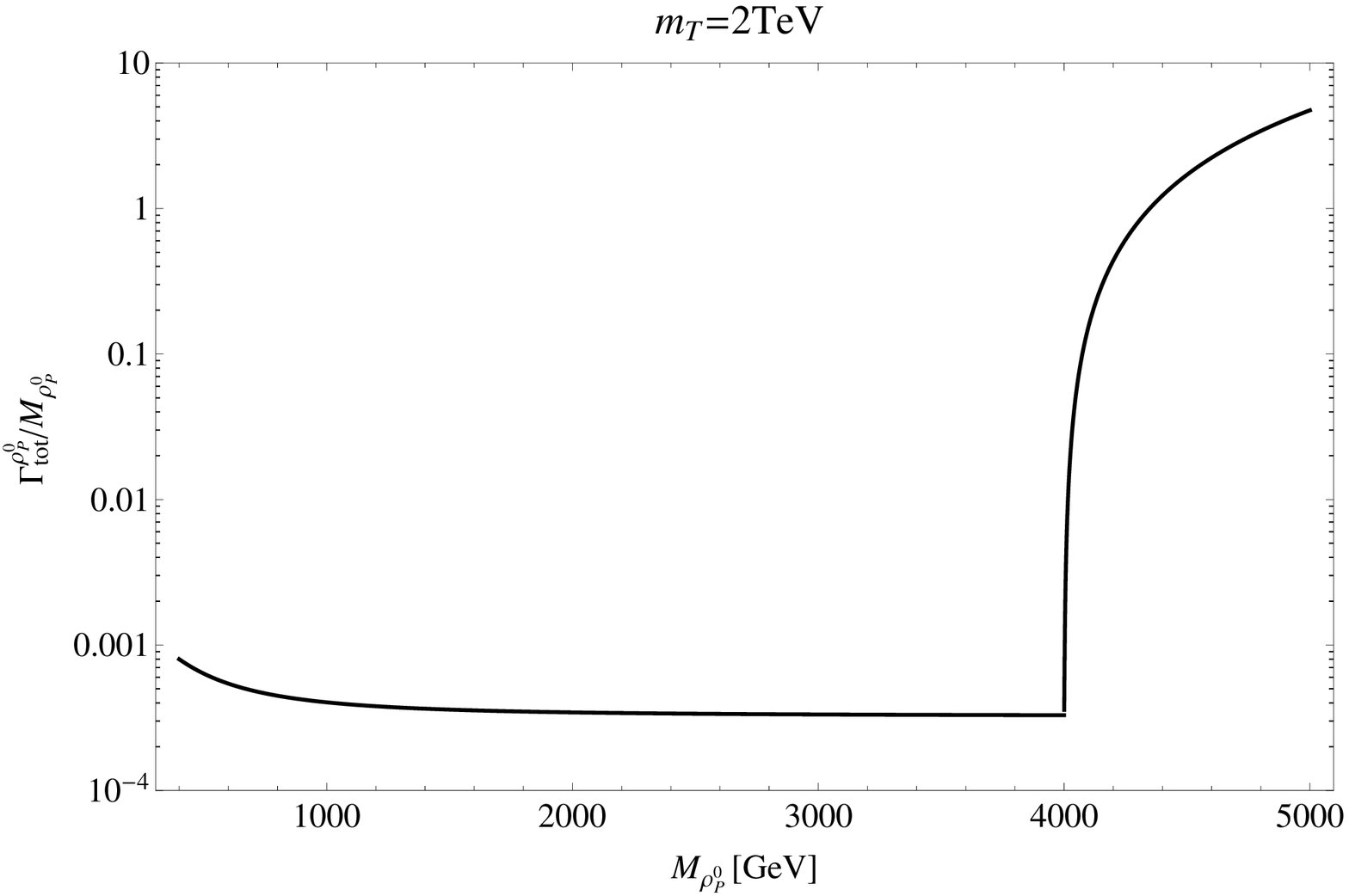}
\caption{ The total widths (normalized by each mass) of the isosinglet technirho mesons, $\rho_{\theta}^0$ (left panel) and $\rho_P^0$ (right panel), as functions of respective technirho masses. Here, we took the masses of the relevant technipions, $M_{T^{3,0,\pm}}$, to be $2$ TeV. 
\label{fig:Twidth1}
}
\end{center} 
 \end{figure}

\begin{figure}[ht]
\begin{center}
   \includegraphics[width=8.5cm]{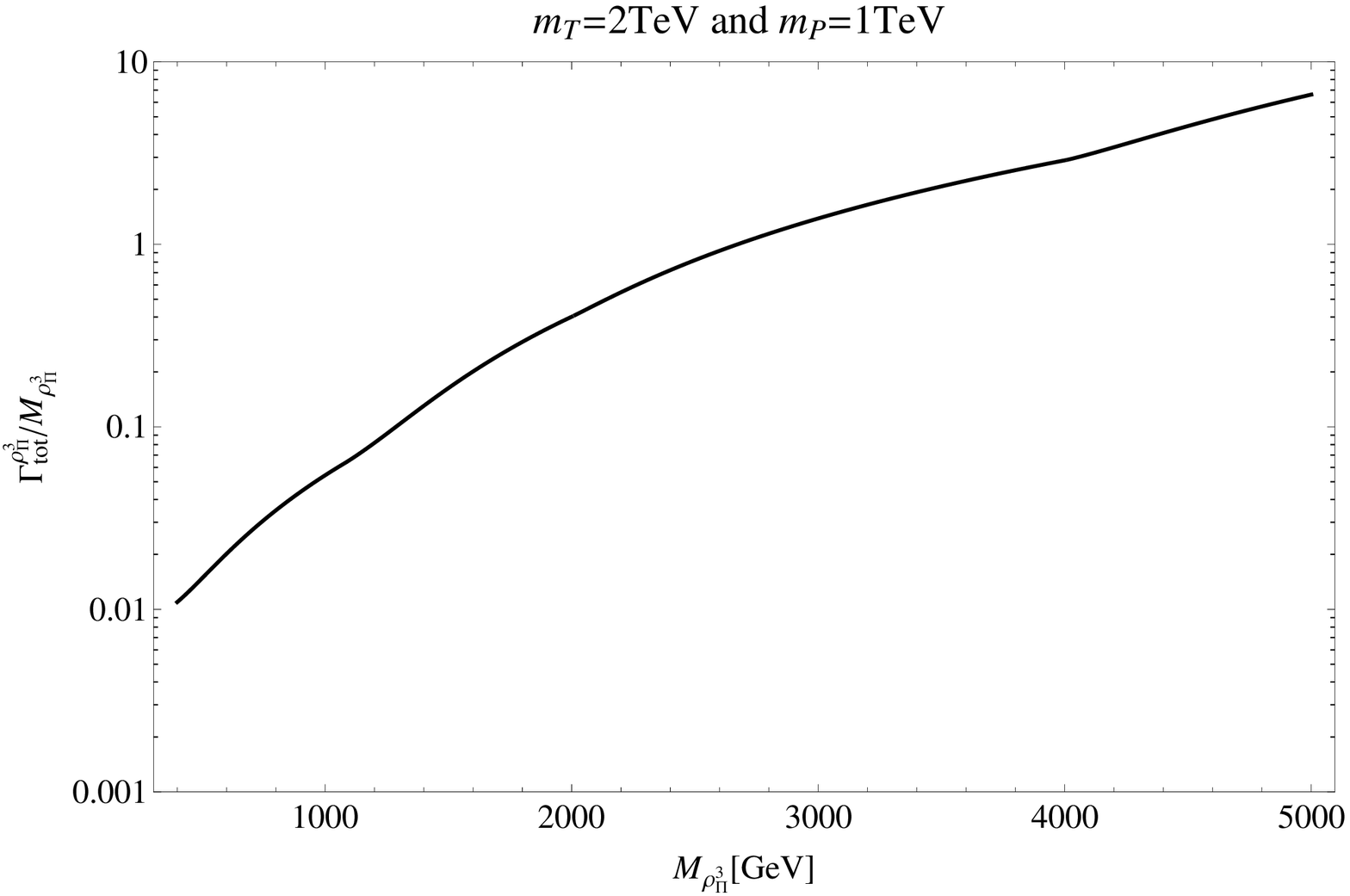} \ \ \ 
   \includegraphics[width=8.5cm]{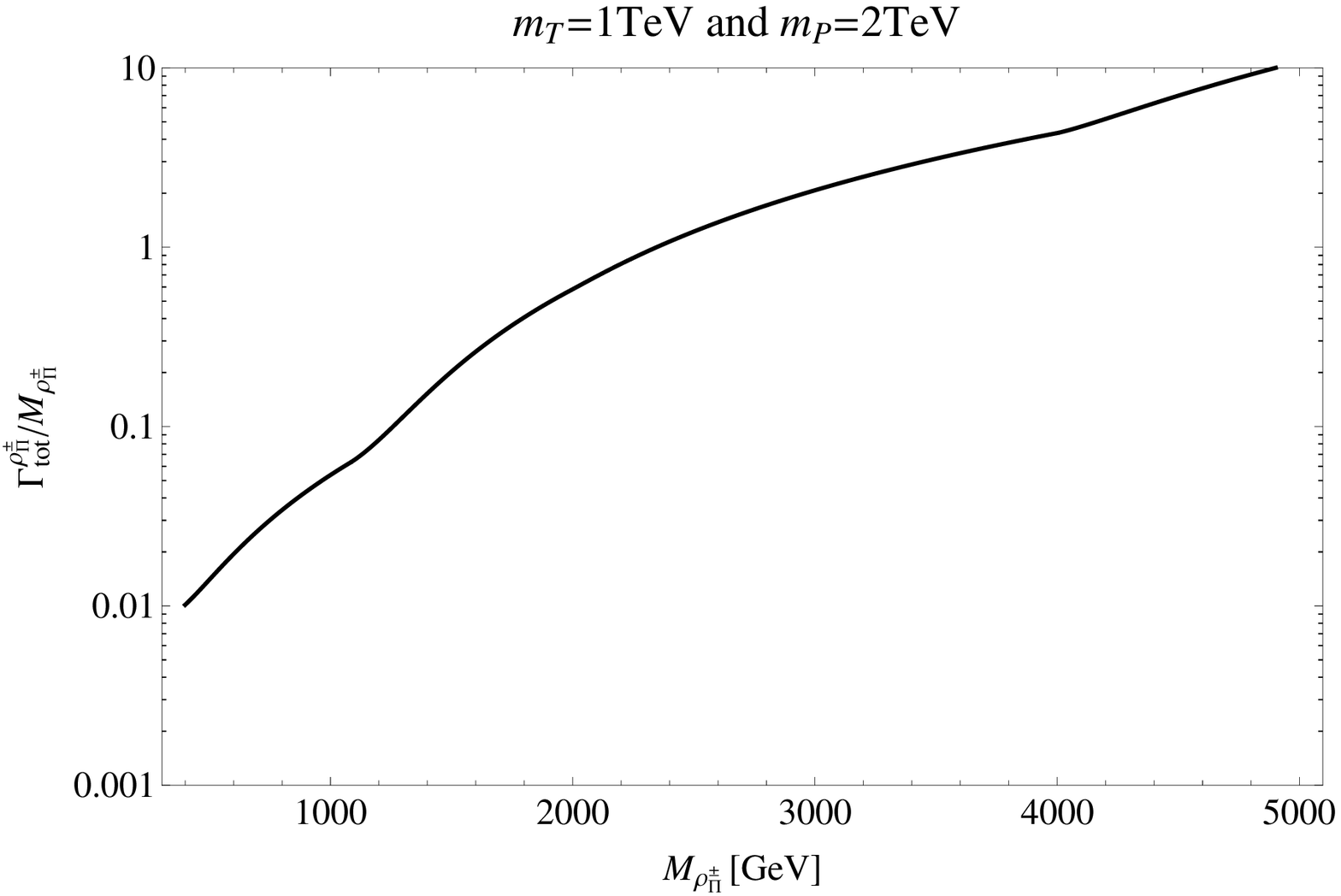}
\caption{ The total widths (normalized by each mass) of the isotriplet technirho mesons, $\rho^3_\Pi$ (left panel) and $\rho^\pm_\Pi$ (right panel), as functions of respective technirho masses. Here, we took the masses of the relevant technipions as  $(M_{T^{3,0,\pm}}, M_{P^{\pm,3}}) = $(2 TeV, 1 TeV). }
\label{fig:Twidth2}
\end{center} 
 \end{figure}

\begin{figure}[ht]
\begin{center}
   \includegraphics[width=8.5cm]{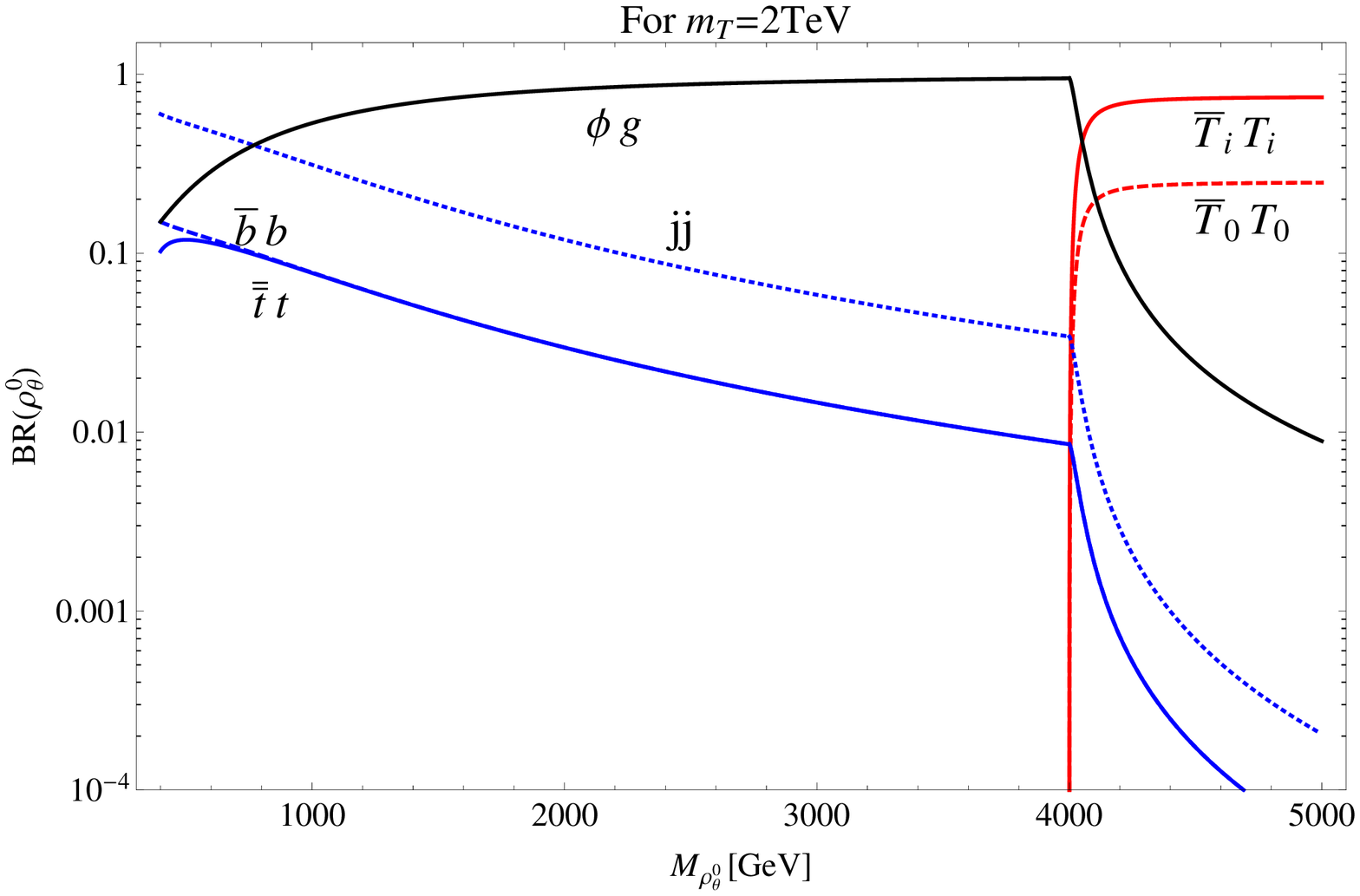}\ \ \ 
   \includegraphics[width=8.5cm]{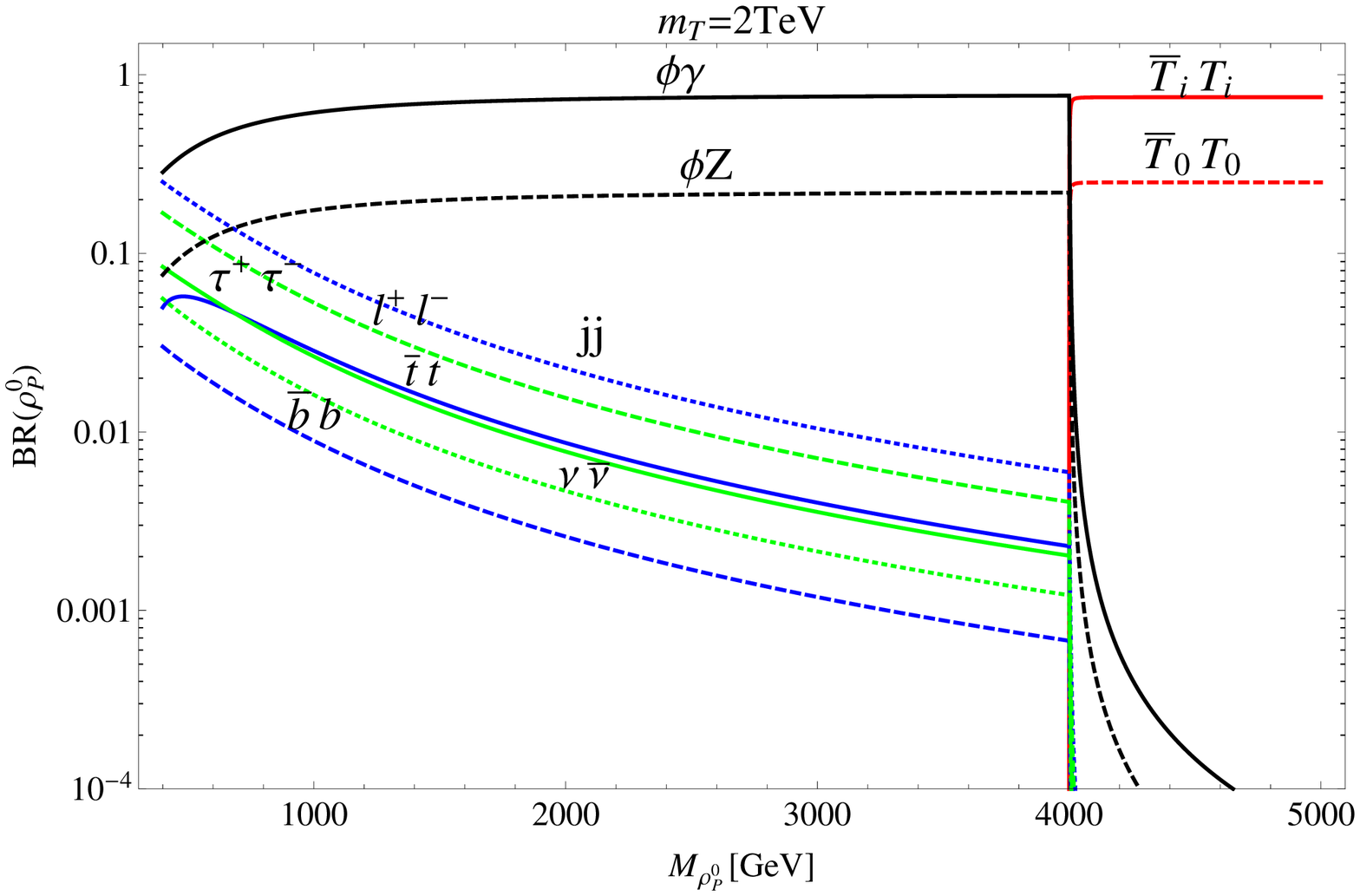}
\caption{ The branching ratios of the isosinglet technirho mesons, $\rho_{\theta}^0$ (left panel) and $\rho_P^0$ (right panel), as functions of respective technirho masses. Here, we took masses of the relevant technipions, $M_{T^{3,0,\pm}}$, to be $2$ TeV. 
Note that $j$ and $l$ in the figure represent the sum of light-quarks $(j= u,d,s,c)$ and that of leptons  
$l=(e, \mu)$, respectively. 
\label{fig:Br1}
}
\end{center} 
 \end{figure}

\begin{figure}[ht]
\begin{center}
   \includegraphics[width=8.5cm]{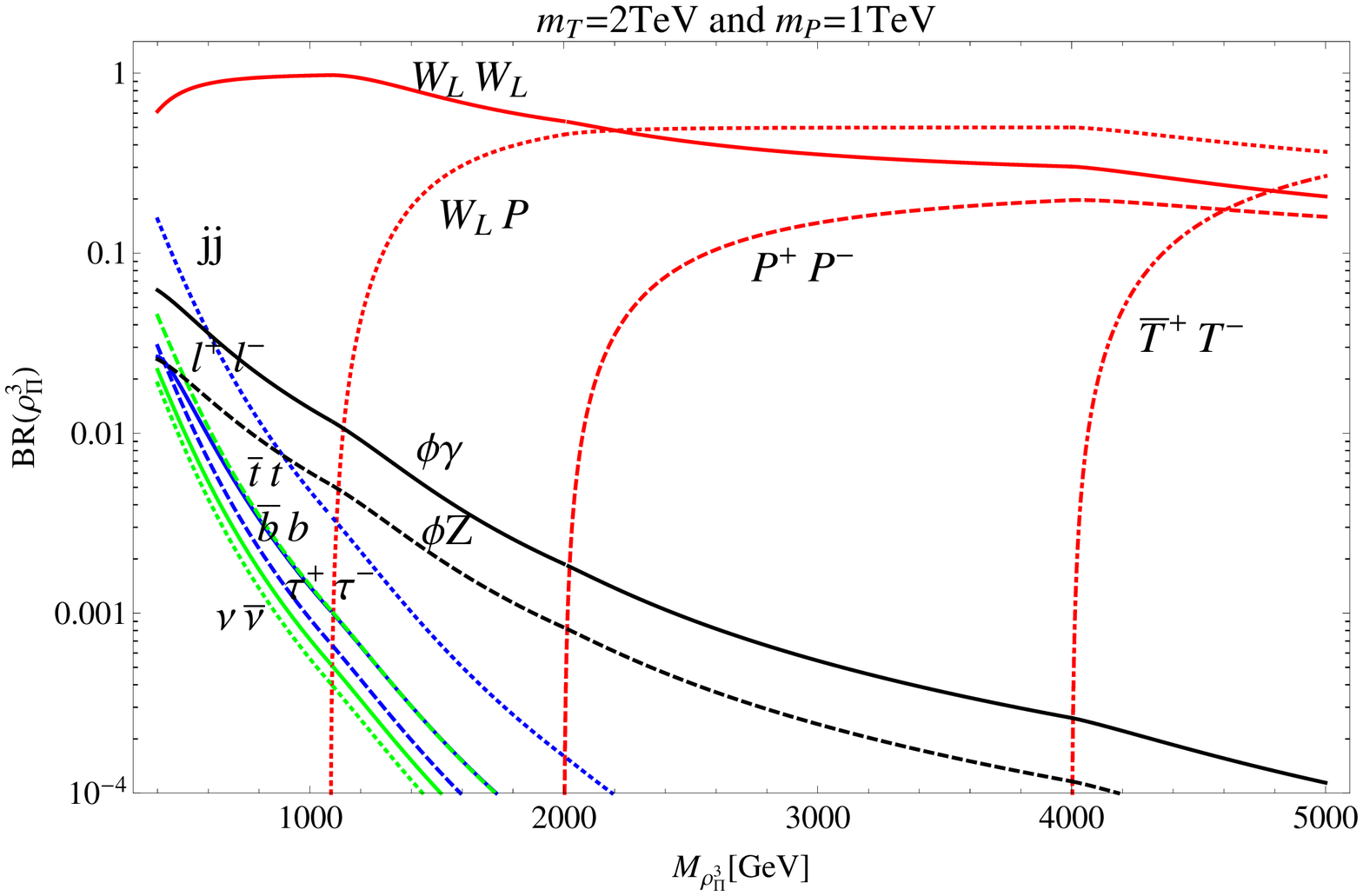}\ \ \ 
   \includegraphics[width=8.5cm]{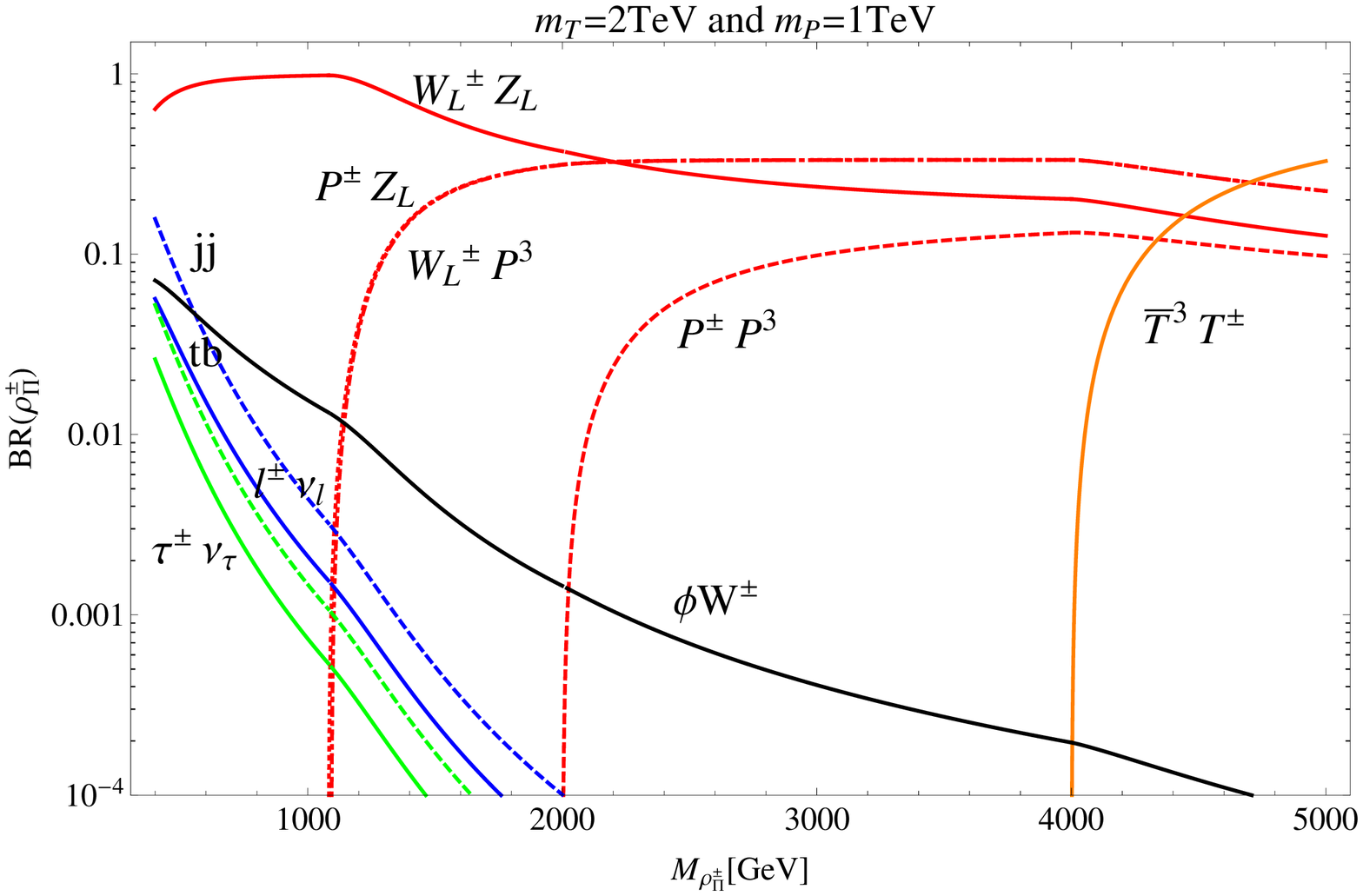}
\caption{ The branching ratios of the isotriplet technirho mesons, $\rho^3_\Pi$ (left panel) and $\rho^\pm_\Pi$ (right panel), as functions of respective technirho masses. Here, we took the masses of the relevant technipions as  $(M_{T^{3,0,\pm}}, M_{P^{\pm,3}}) = $(2 TeV, 1 TeV). 
Note that $j$ and $l$ in the figure represent the sum of light-quarks $(j= u,d,s,c)$ and that of leptons  
$l=(e, \mu)$, respectively. 
\label{fig:Br2}
}
\end{center} 
\end{figure}

Partial decay rates of technirho mesons are calculated by using relevant vertex terms summarized in Appendix~\ref{app:int}, and we show the explicit expressions of them for $\rho_{\theta}^0$, $\rho_P^0$ and $\rho^{\pm ,3}_\Pi$ in Appendix~\ref{app:Pdecay}. The total decay widths are plotted in Figs.~\ref{fig:Twidth1} and \ref{fig:Twidth2} as functions of respective technirho masses.
The branching fractions for $\rho_{\theta}^0, \rho_P^0$ and $\rho^{\pm, 3}_\Pi$ are shown in Figs.~\ref{fig:Br1} and \ref{fig:Br2}, respectively. 
It should also be noted~\cite{Kurachi:2014xla} that the technipion masses are severely constrained by the LHC data to be of the order of TeV~\footnote{
The perturbative treatment of such a large explicit breaking effect might sound questionable. Actually, this is a typical phenomenon of the 
``amplification of a symmetry violation'' by the large anomalous dimension in the dynamics near the criticality, resulting in a huge violation effects for small violation parameter, as
 was most dramatically shown in the top quark condensate model~\cite{Miransky:1988xi}. 
See e.g., the first reference in  \cite{Yamawaki:1996vr}.
One also might suspect that  since all the massive pseudo NG bosons are decoupled, leaving only three exact NG bosons absorbed into
the $W/Z$ bosons, 
the theory would be equivalent to the model
based on the $G/H=SU(2)_L \times SU(2)_R/SU(2)_V$, the one-doublet model.  
However,  
the fictitious NG bosons (absorbed into $W/Z$) as well as the TD are composites of all the 8 flavors 
technifermions, not a particular subset of them, which contribute to the dynamics on the same footing. Thus the walking dynamics responsible for the lightness of the TD as well as the FCNC solution is still operative in contrast to the one-doublet model. 
}. 
Here we take reference values of the masses of technipions relevant to the calculations as $(M_{T^{3,0,\pm}}, M_{P^{\pm,3}}) = $(2 TeV, 1 TeV). 
This choice is motivated by the results of Refs.~\cite{Jia:2012kd, Kurachi:2014xla}, 
in which it was shown that color-triplet technipions are heavier than the color-singlet technipions, and can be as heavy as $O(1)$ TeV. Changing the reference values of $M_{T^{3,0,\pm}}$ and $M_{P^{\pm,3}}$ does not affect the LHC analysis in this paper as will be shown later. 
From these figures, we see the following general tendency: technirhos dominantly decay to $\pi \pi$ or $\pi W_L (\pi Z_L)$
above the thresholds (besides $W_L W_L$/$W_L Z_L$ channels for $\rho_{\Pi}^{\pm, 3}$), 
and decay widths become large; 
on the other hand, below these thresholds, the decay widths are small enough that usual resonance search strategies can be applied. 
In particular, the $\rho_{\theta}^0$ ($\rho_P^0$) search below the technipion-pair-decay threshold is interesting in the sense that it dominantly decays to the TD and $g$ ($\gamma/Z$). 
The associate production of the Higgs and the weak gauge boson ($W, Z$) from the resonant vector meson were discussed in the literature in the context of certain kinds of dynamical EW symmetry-breaking scenarios~\cite{Zerwekh:2005wh, Belyaev:2008yj, Castillo-Felisola:2013jua}. On the other hand, the associated production of the Higgs (TD) and the gluon from the color-octet vector resonance is a characteristic process of the one-family model, which plays an important and complementary role together with existing studies of color-octet signals (see, e.g., Ref.~\cite{Atre:2013mja}).

\section{Technirho phenomenology at the LHC} 
\label{sec:pheno}

As mentioned in the previous section, we consider the production of the technirho mesons through the mixing with the SM gauge bosons which are produced by the Drell-Yan process (Fig.~\ref{fig:DY}). The LHC cross section of technirho mesons with mass $M_\rho$ is thus calculated to be (for a review, see, e.g., Ref.~\cite{Chivukula:1995dt}) 
\begin{eqnarray} 
  \frac{d \sigma (pp \to \rho X \to AB \, X) }{d \eta \, d M^2} 
 &= &
 \frac{32 \pi}{s}   
\, \sum_{a,b} C_{ab} \, f_{a/p}\left(\frac{M}{\sqrt{s}} e^\eta, M^2 \right) 
f_{b/p}\left( \frac{M}{\sqrt{s}} e^{-\eta},M^2 \right)  
\nonumber \\ 
&&
\times  
\frac{C_\rho \cdot (2 S_\rho + 1)}{(2S_a + 1)(2 S_b + 1)} \frac{M^2}{M_\rho^2} 
\frac{\Gamma(\rho \to ab) \Gamma(\rho \to AB)}{(M^2 - M_\rho^2)^2+M_\rho^2 (\Gamma^{\rho}_{\rm tot})^2} 
\,, \label{cross-formula}
\end{eqnarray}
where the function $f_{a/p}$ denotes the parton distribution function for parton $a$ in the proton, which is available from Ref.~\cite{Pumplin:2002vw} (for CTEQ6M); 
$\Gamma_{\rm tot}^{\rho}$ is the total width of $\rho$; 
$\sqrt{s}$ is the center-of-mass energy at the LHC ($\sqrt{s}=8$ or 14 TeV); 
$\eta$ is the rapidity of the $a$-$b$ system in the $p$-$p$ center-of-mass frame; $C_{ab}$ is a multiplication factor regarding the $SU(3)$ gauge group [e.g., $C_{gg}=(1/8)^2, C_{qq}=(1/3)^2$]; 
$M^2$ is an invariant mass squared associated with particles $A$ and $B$ coming out of the $\rho$; 
$C_\rho=1(8)$ for the color-singlet (-octet) technirho meson;  
and $(2 S_a +1)$ is a multiplication factor 
for spin degeneracy [e.g., $(2S_a+1)=2$ for $a=q$, and $(2 S_\rho +1) = 3$ for vector mesons].  

\begin{figure}[t]
\begin{center}
   \includegraphics[width=8.5cm]{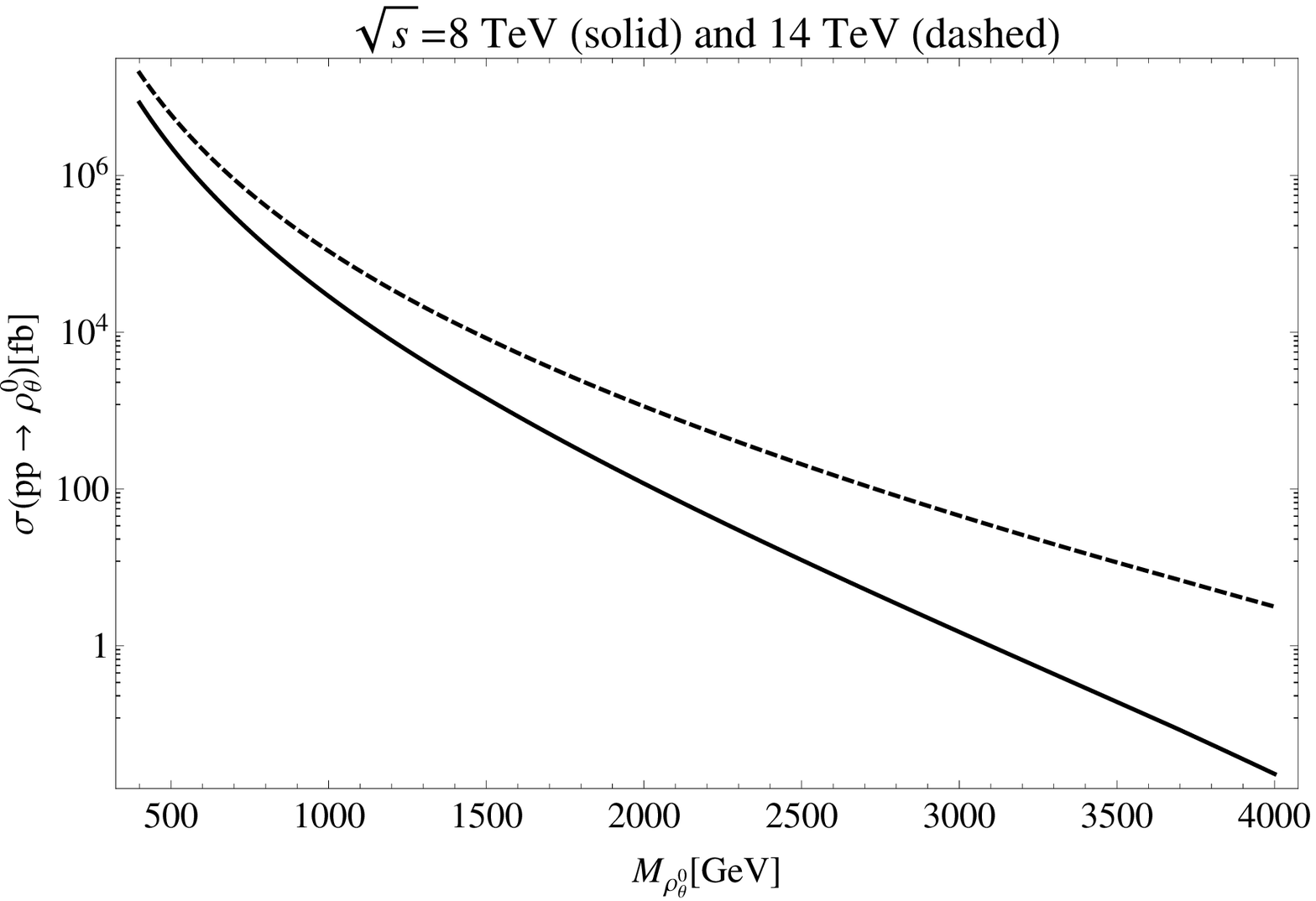}\ \ \ 
   \includegraphics[width=8.5cm]{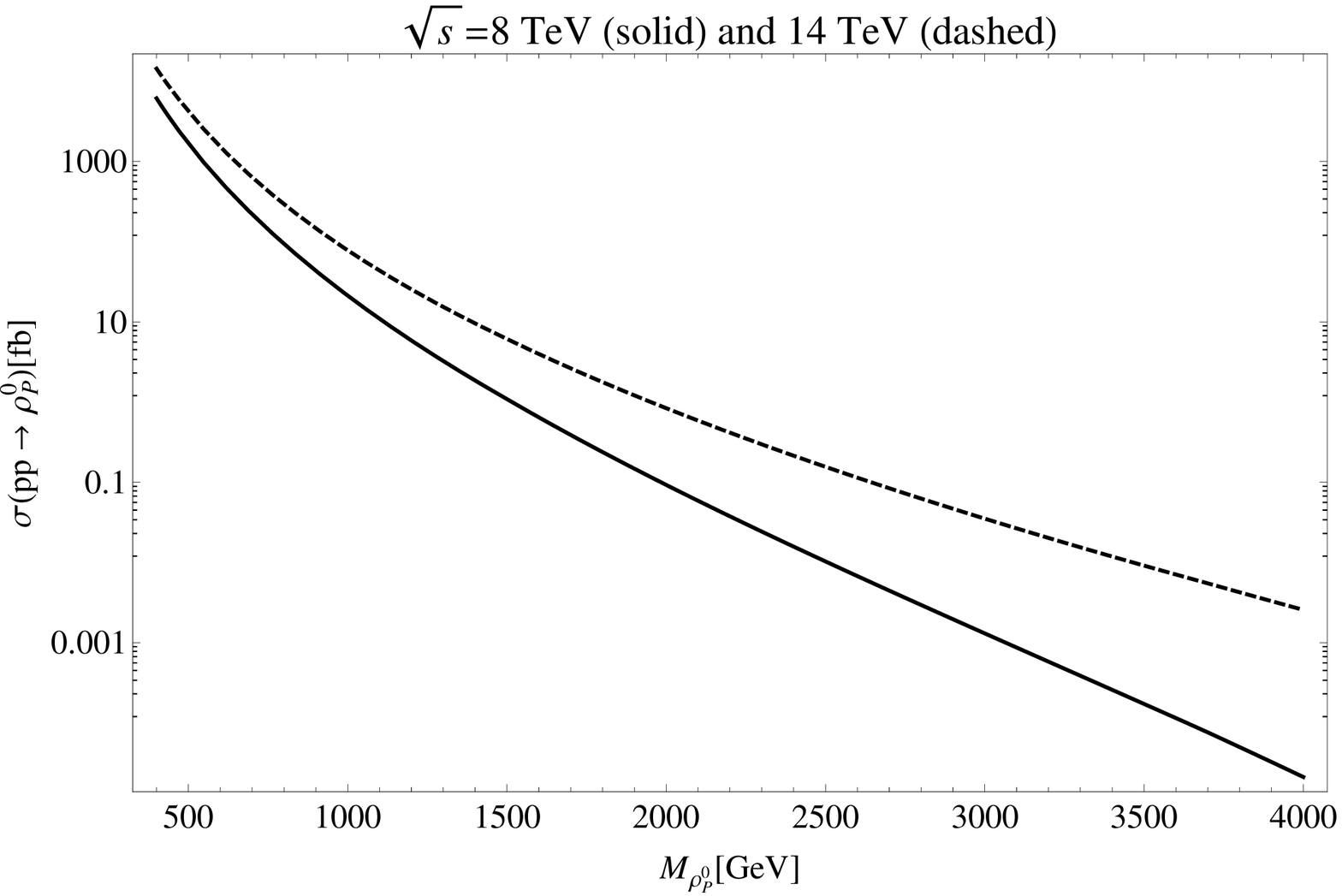}
\caption{ The LHC production cross section of the isosinglet technirho mesons, $\rho_{\theta}^0$ (left panel) and $\rho_P^0$ (right panel), as functions of respective technirho masses. Here, we took the masses of the relevant technipions as  $(M_{T^{3,0,\pm}}, M_{P^{\pm,3}}) = $(2 TeV, 1 TeV).
The solid and dashed curves correspond to $\sqrt{s}=$ 8 and 14 TeV, respectively. 
 }
\label{fig:prod1}
\end{center} 
 \end{figure}

\begin{figure}[t]
\begin{center}
   \includegraphics[width=8.5cm]{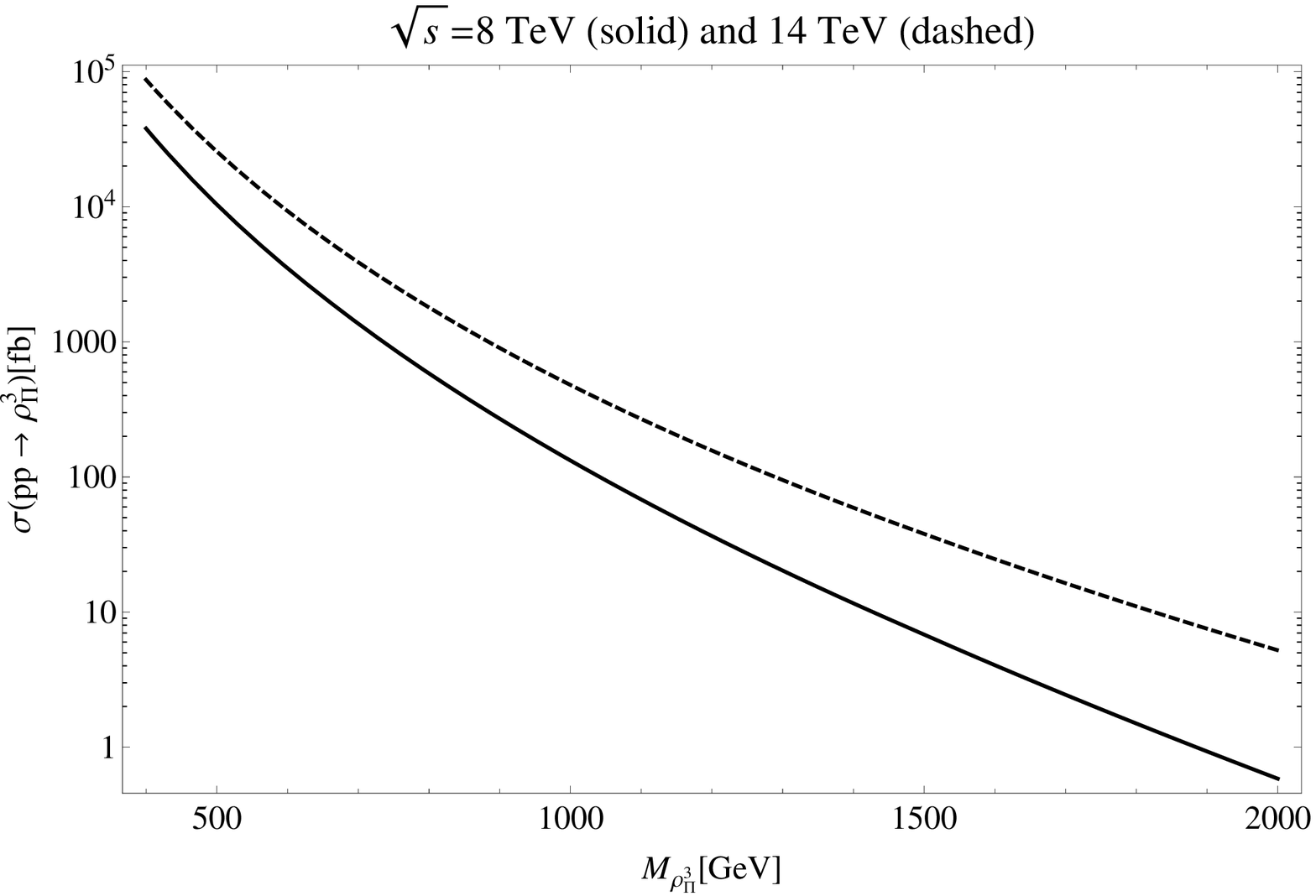}\ \ \ 
   \includegraphics[width=8.5cm]{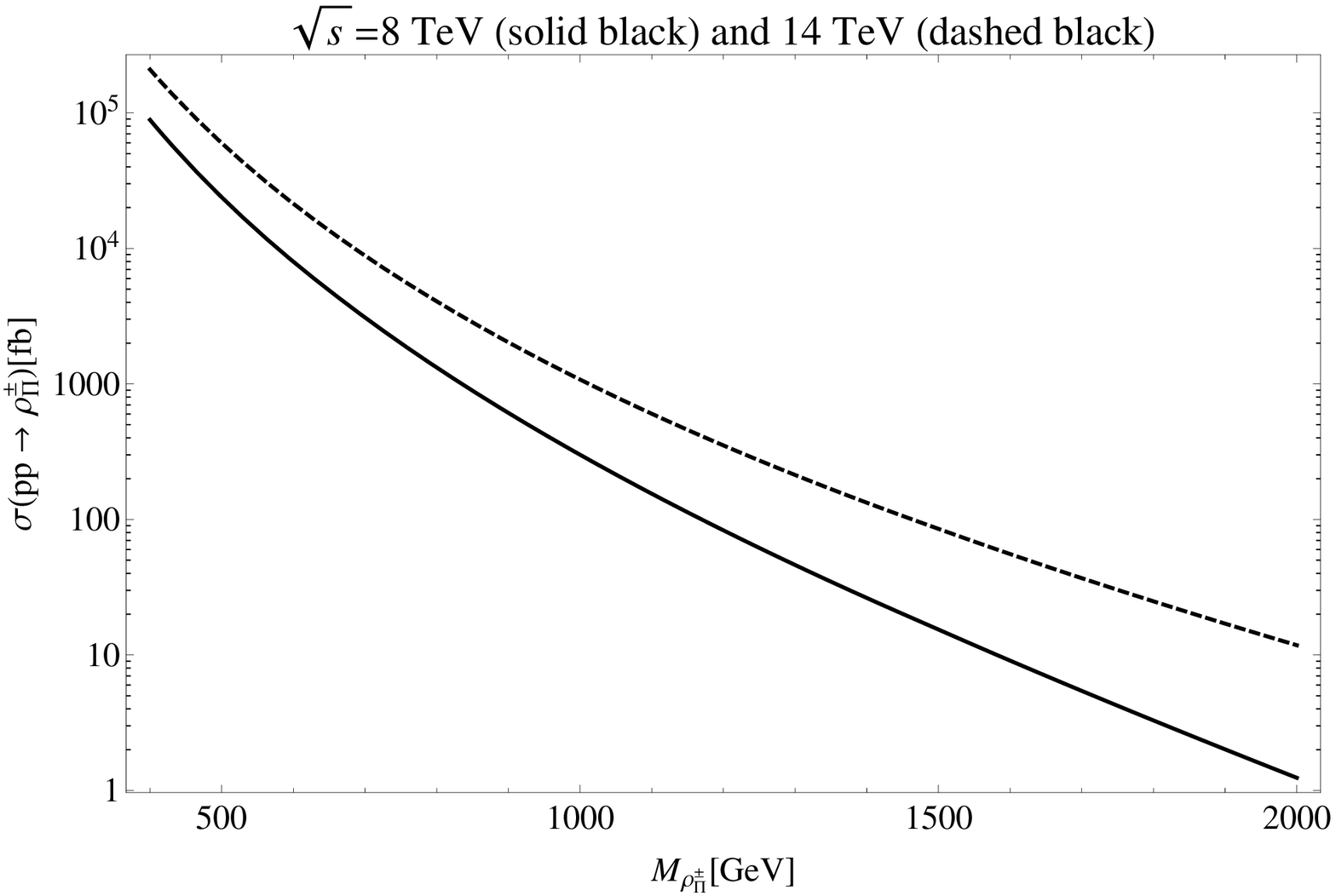}
\caption{The LHC production cross section of the isotriplet technirho mesons, $\rho^3_\Pi$ (left panel) and $\rho^\pm_\Pi$ (right panel), as functions of respective technirho masses. 
Here, we took the mass of the relevant technipion, $M_{T^{3,0,\pm}}$, to be $2$ TeV. 
The solid and dashed curves correspond to $\sqrt{s}=$ 8 and 14 TeV, respectively. } 
\label{fig:prod2}
\end{center} 
 \end{figure}

Figures~\ref{fig:Twidth1} and \ref{fig:Twidth2} show that 
$\Gamma_{\rm tot}^\rho/M_\rho \ll 1$ when $M_\rho < 4$ TeV for $\rho_{\theta}^0, \rho_P^0$ and $M_\rho < 2$ TeV for $\rho^{3,\pm}_\Pi$, 
so we may apply the narrow-width approximation when evaluating Eq.(\ref{cross-formula}) by replacing 
the $\rho$-resonance function $1/[(M^2 - M_\rho^2)^2 + M_\rho^2 (\Gamma_{\rm tot}^\rho)^2]$ 
with $\pi/(M_\rho \Gamma_{\rm tot}^{\rho}) \delta(M^2 - M_\rho^2)$. 
For the Drell-Yan production $(\rho=\rho_{\theta}^0, \rho^{\pm, 3}_\Pi, \rho_P^0)$, 
we thus have 
\begin{eqnarray} 
   \sigma_{\rm DY} (pp \to \rho \to AB)  
 &= &
 \frac{16 \pi^2}{3 s} C_\rho   
  \frac{{\rm BR}(\rho \to AB)}{M_{\rho}} 
  \nonumber \\ 
  && 
  \times 
\, \sum_{q={\rm quarks}} \Gamma(\rho \to q\bar{q})\, 
\int_{-Y_B}^{Y_B} d\eta \, f_{q/p}\left(\frac{M_{\rho}}{\sqrt{s}} e^\eta, M_{\rho}^2 \right) 
f_{\bar{q}/p}\left( \frac{M_{\rho}}{\sqrt{s}} e^{-\eta},M_{\rho}^2 \right)  \,,
\end{eqnarray}
where $Y_B = -\frac{1}{2}\ln(M_\rho^2/s)$.
The predicted production cross sections of $\rho_{\theta}^0$, $\rho_P^0$, and $\rho^{3,\pm}_\Pi$ are plotted in Figs.~\ref{fig:prod1} and \ref{fig:prod2}, respectively.

\subsection{Current LHC limits}

The 8 TeV LHC data have already placed stringent constraints on masses of hypothetical 
heavy resonances, such as $Z'$, $W'$, colorons, Kaluza-Klein (KK) gluons, etc. 
Here, by using these results, we give a rough estimate of the lower bound on the  masses of technirho mesons ($\rho_{\theta}^0, \rho_P^0$ and $\rho^{\pm,3}_\Pi$) in the one-family model under the assumption that the kinematics of the technirho production and decay process is more or less the same as that of, e.g., $Z'$ and $W'$.

The resonance search in the dijet mass distribution by CMS~\cite{CMS8-jj} places the strongest constraint on the $\rho_{\theta}^0$ mass, while the dilepton resonance search by ATLAS~\cite{ATLAS82l} and CMS~\cite{CMS82l} experiments are the most relevant for $\rho_P^0$ and $\rho^3_\Pi$. As for $\rho^\pm_\Pi$, studies of resonant $WZ \rightarrow 3 \ell + \nu$ production by ATLAS~\cite{ATLAS8WZ} and CMS~\cite{CMS8WZ} experiments place the strongest constraint on its mass. The $t \bar{t}$ resonance search 
by ATLAS~\cite{ATLAS8WZ} and CMS~\cite{CMS8WZ} also places a strong constraint on the $\rho_{\theta}^0$ mass, though the bound is slightly weaker than that obtained from the dijet search. $\rho_P^0$ and $\rho^3_\Pi$ also decay to $\bar{t} t$, though the cross sections are so small that the current LHC data do not give any constraint.

\begin{figure}[t]
\begin{center}
   \includegraphics[width=8.5cm]{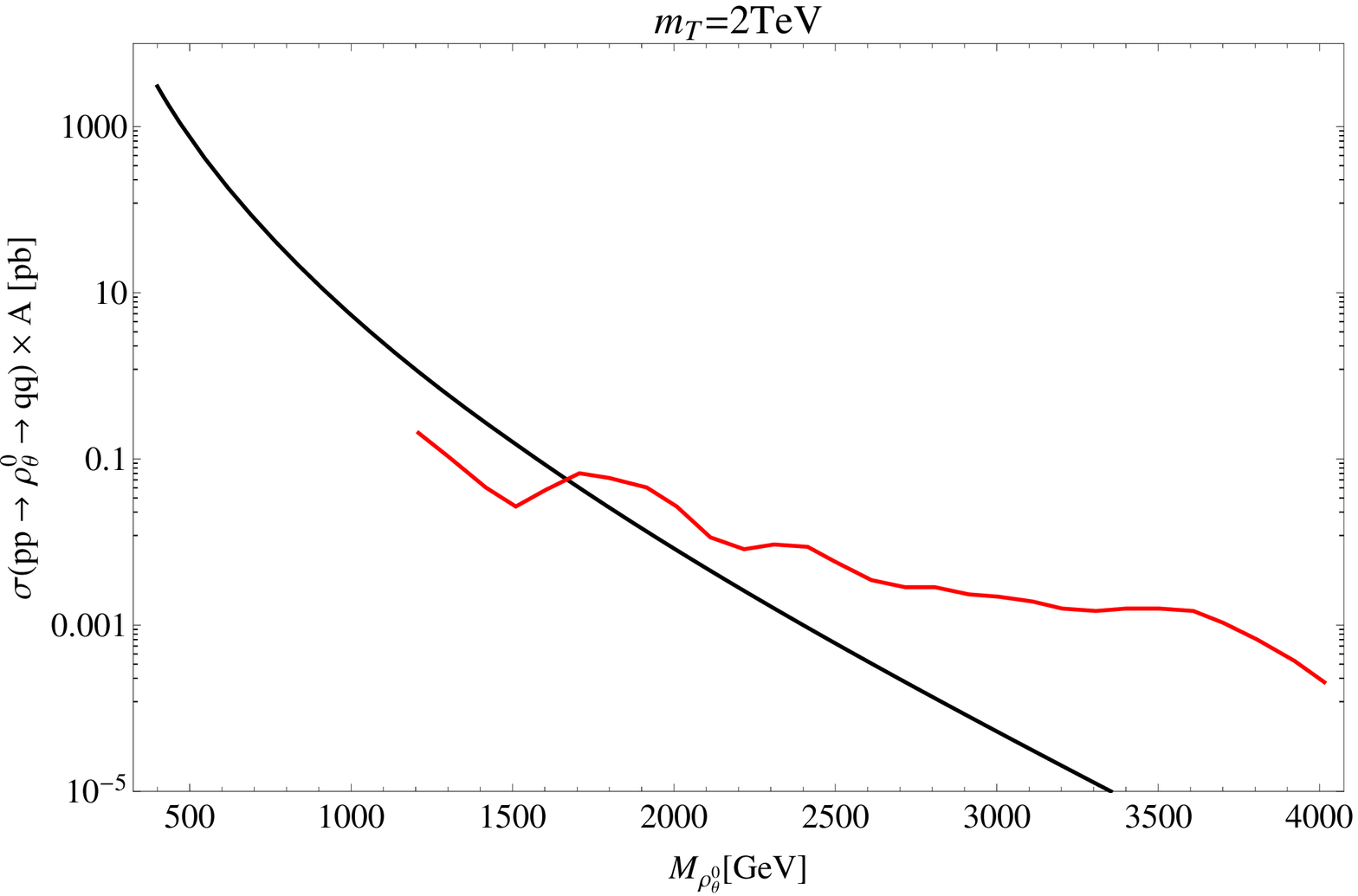}\ \ \ 
   \includegraphics[width=8.5cm]{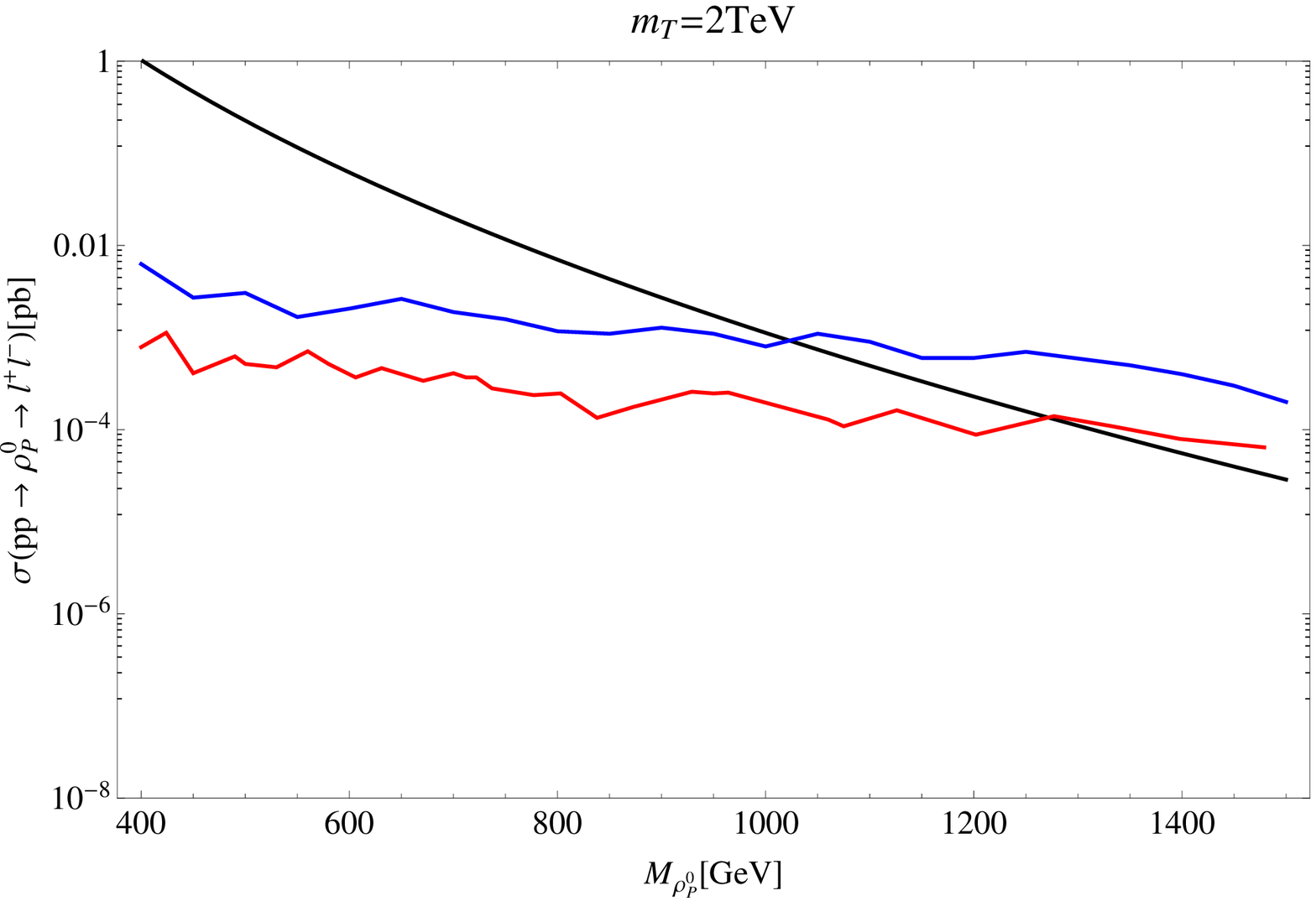}
\caption{{\bf Left panel:}   
$\sigma_{\rm DY}(pp \to \rho_{\theta}^0 \to qq)$, multiplied by acceptance $A \simeq 0.6$~\cite{Khachatryan:2010jd,CMS8-jj}, for $q=u,d,s,c$,  
in unit of pb as a function of $M_{\rho_{\theta}^0}$.
The black curves correspond to prediction of the one-family model for $\sqrt{s}=$ 8 TeV. The 95\% C.L. upper limit 
on a generic cross section times acceptance set by searches for a new resonance in dijet mass distribution 
by CMS experiment with $\sqrt{s}=$8 TeV ~\cite{CMS8-jj} is shown as red curve.
{\bf Right panel:} $\sigma_{\rm DY}(pp \to \rho_P^0 \to l^+l^-)$ with $l=e, \mu$ 
as a function of $M_{\rho_P^0}$. The black curve corresponds to the prediction of the one-family model for $\sqrt{s}=$ 8 TeV.   
The 95\% C.L. upper limits on $Z'\to l^+ l^-$ cross section reported by  ATLAS~\cite{ATLAS82l} and CMS~\cite{CMS82l} experiments have been 
drown by the blue and red curves, respectively. 
\label{fig:const1}}
\end{center} 
 \end{figure} 
\begin{figure}[t]
\begin{center}
   \includegraphics[width=8.5cm]{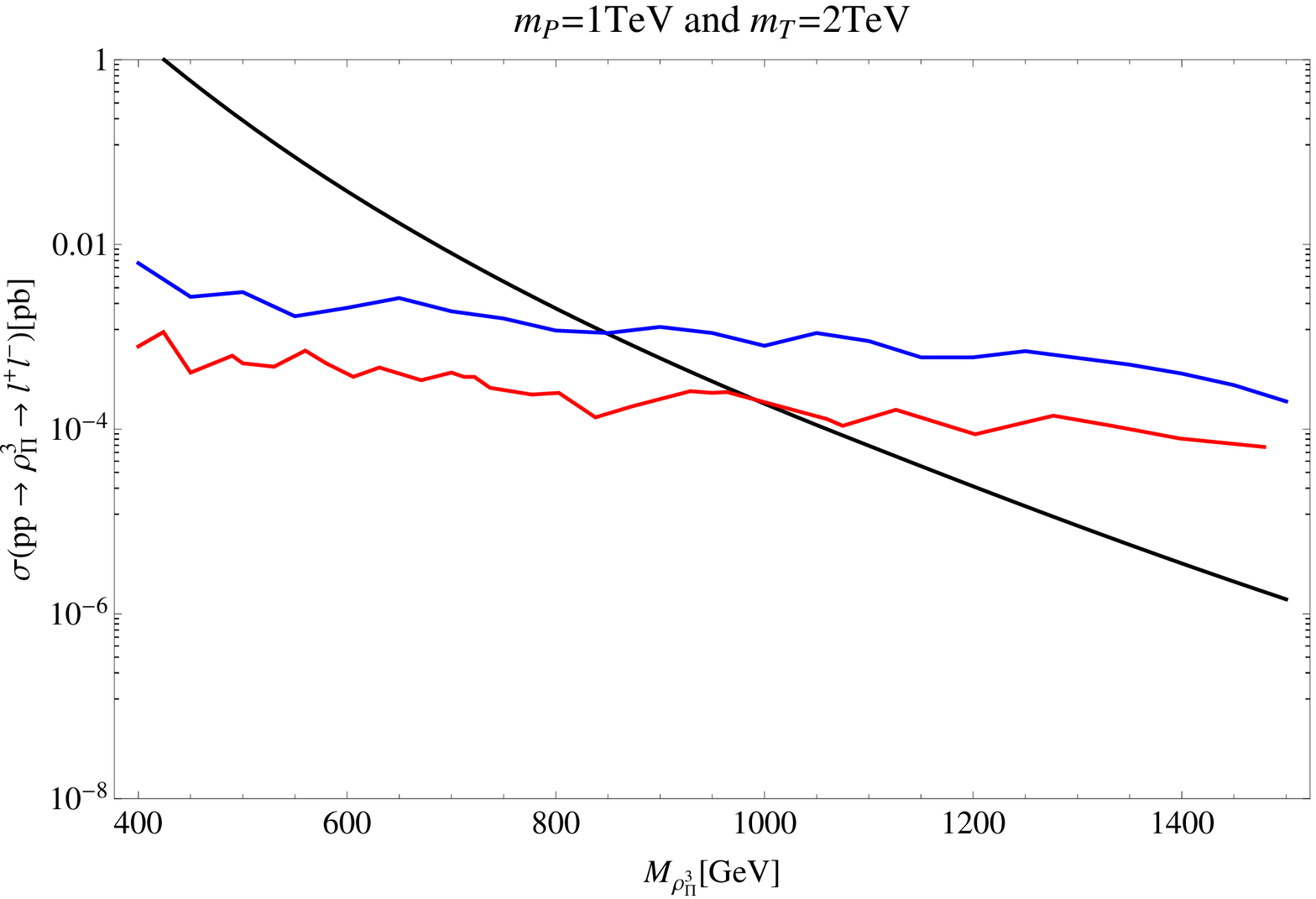}\ \ \ 
   \includegraphics[width=8.5cm]{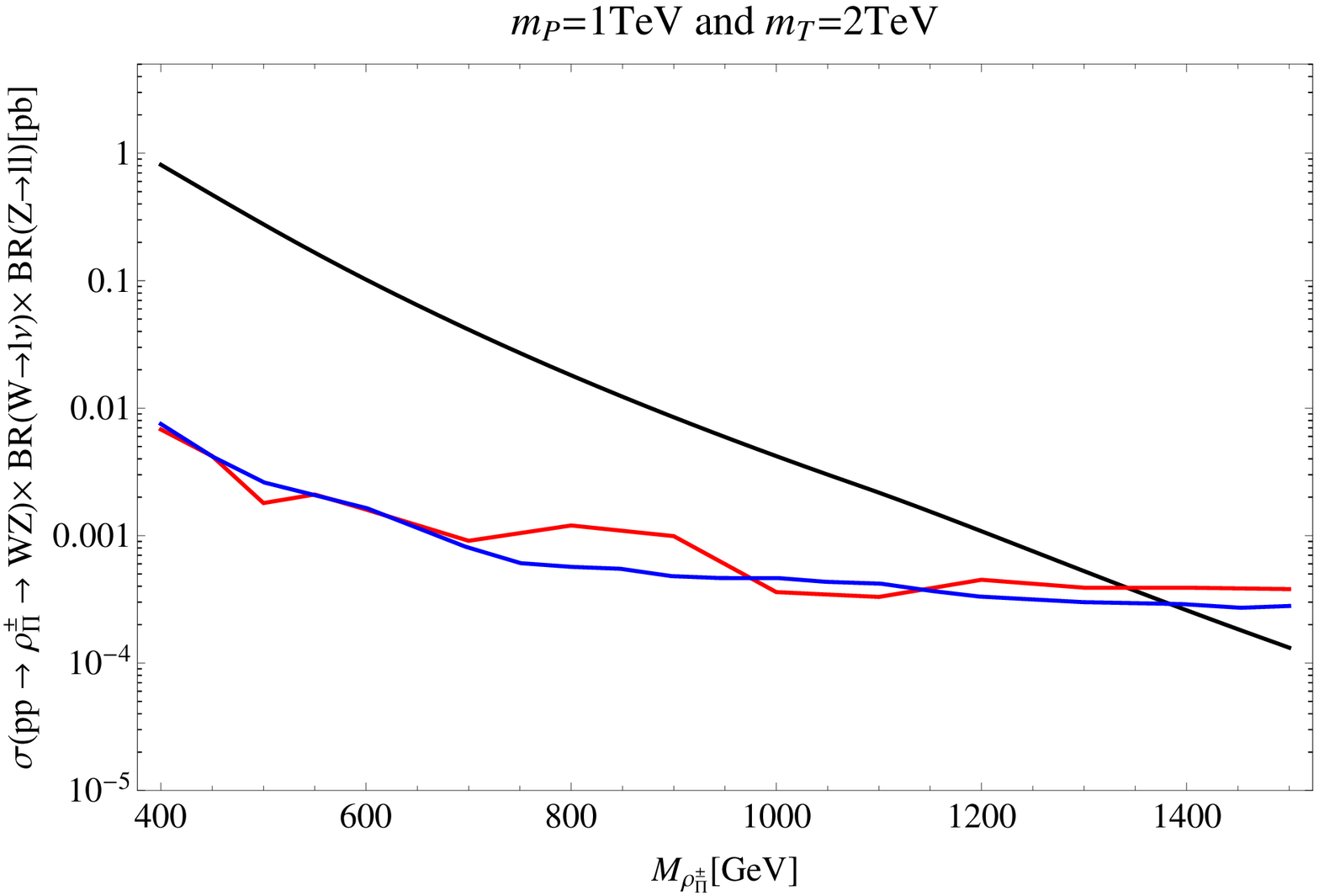}
\caption{{\bf Left panel:} $\sigma_{\rm DY}(pp \to \rho^3_\Pi \to l^+l^-)$ with $l=e,\mu$ as a function of $M_{\rho^3_\Pi}$. The black curve corresponds to the prediction of the one-family model for $\sqrt{s}=$ 8 TeV.    
The 95\% C.L. upper limits on $Z'\to l^+ l^-$ cross section reported by  ATLAS~\cite{ATLAS82l} and CMS~\cite{CMS82l} experiments have been 
drown by the blue and red curves, respectively. 
{\bf Right panel:}
$\sigma_{\rm DY}(pp \to \rho^\pm_\Pi \to W^\pm Z) \times {\rm BR}(W^\pm \to l^\pm \nu ) \times {\rm BR}(Z \to l^+ l^-)$ with $l=e,\mu$. 
The black curve corresponds to the prediction of the one-family model for $\sqrt{s}=$ 8 TeV.  
The 95\% C.L. upper limits on $W'/\rho \to WZ$ cross section reported by  ATLAS~\cite{ATLAS8WZ} and CMS~\cite{CMS8WZ} experiments have been 
drown by the blue and red curves, respectively.
\label{fig:const2}}
\end{center} 
\end{figure} 
In Figs.~\ref{fig:const1} and \ref{fig:const2}, we show the cross sections for isosinglet rho mesons ($\rho_{\theta}^0, \rho_P^0$) and isotriplet rho mesons 
($\rho^3_\Pi, \rho^\pm_\Pi$), respectively, for the most constrained decay processes mentioned above together with the experimental upper bounds.
Here, we take the technipion masses as $M_{T^{\pm,3}}$ = 2 TeV and $M_{P^{\pm, 3}}$ = 1 TeV, so that all the relevant energy regions are below the threshold of the decay channels that involve technipion(s). 
From these figures, we find that the current LHC experiments constrain the masses of the technirho mesons to be 
\begin{eqnarray} 
  M_{\rho_{\theta}^0} & \gtrsim & 1.7 \, {\rm TeV} 
  \,, \nonumber \\ 
  M_{\rho_P^0} & \gtrsim & 1.3 \, {\rm TeV} 
  \,, \nonumber \\ 
  M_{\rho^3_\Pi} &\gtrsim & 1.0 \, {\rm TeV} 
  \,, \nonumber \\ 
  M_{\rho^\pm_\Pi} &\gtrsim  & 1.4 \, {\rm TeV} 
  \,. \label{limit:mass}
\end{eqnarray} 
As we mentioned in the previous section, these results are insensitive to the precise values of $M_{T^{\pm,3}}$ and $M_{P^{\pm, 3}}$ as long as the relevant mass range of the technirho is below the technipion thresholds. 

Note that the limits on the technirho masses are milder than those on other hypothetical spin-1 resonances such as $W', Z'$, KK gluons, and colorons. 
This is due to the fact that the technirho mesons have no direct couplings to the SM quarks, 
and hence the Drell-Yan productions necessarily arise through the $\rho-{\cal V}$ mixing, as in Eq.(\ref{v-rho:mixing}), leading to the significant suppression by $\alpha_s$ or $\alpha_{\rm em}$ in the amplitudes compared to the case for other hypothetical spin-1 resonances.

\subsection{Associated production of the technidilaton and the SM gauge boson through a resonant technirho}

\begin{figure}[t]
\begin{center}
   \includegraphics[width=15cm]{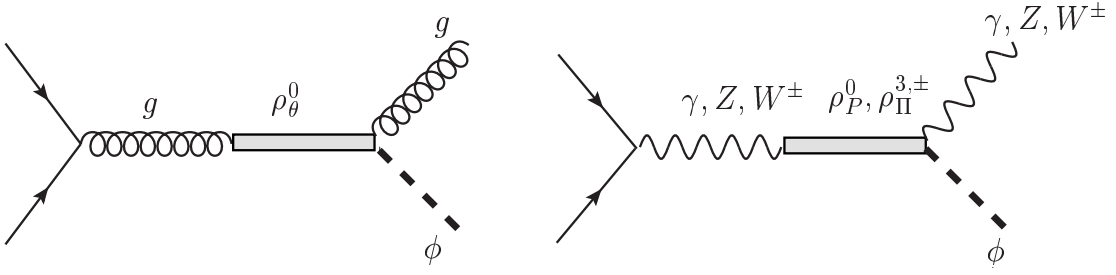}
\caption{DY production of the technirho meson which decays into the Higgs (TD) and the SM gauge boson.
\label{fig:dilaton}}
\end{center} 
\end{figure}

As we mentioned at the end of Sec.~III, the most interesting search channel of the one-family model is the DY production of the technirho meson which decays into the Higgs (TD) and the SM gauge boson (see Fig.~\ref{fig:dilaton}). The processes consist of vertices in Eqs.(\ref{v-rho:mixing:0}) and (\ref{TD-rho-V}) as a consequence of the scale-invariant extension of the HLS formalism.

\begin{figure}[t]
\begin{center}
   \includegraphics[width=8.5cm]{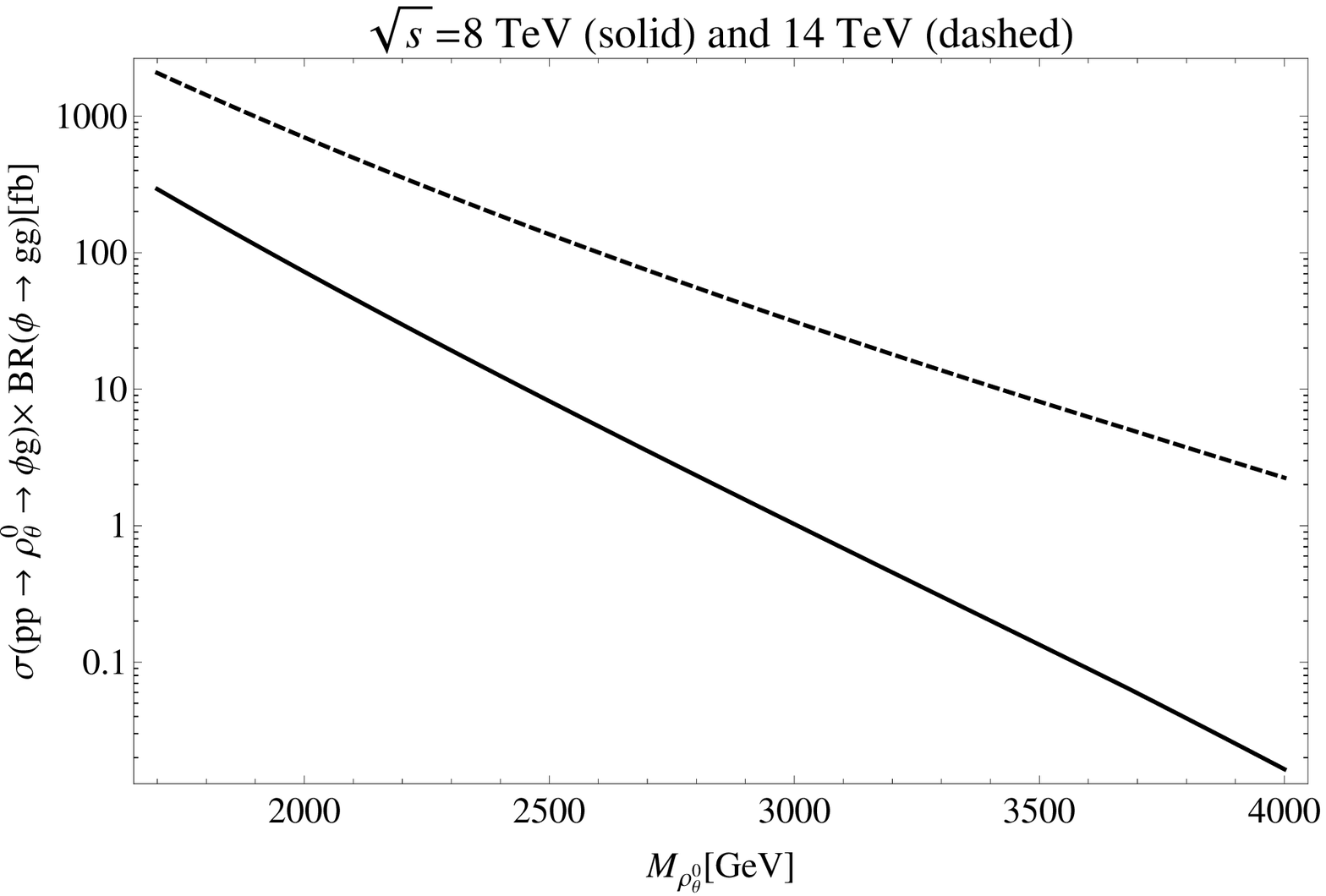}\ \ \ 
   \includegraphics[width=8.5cm]{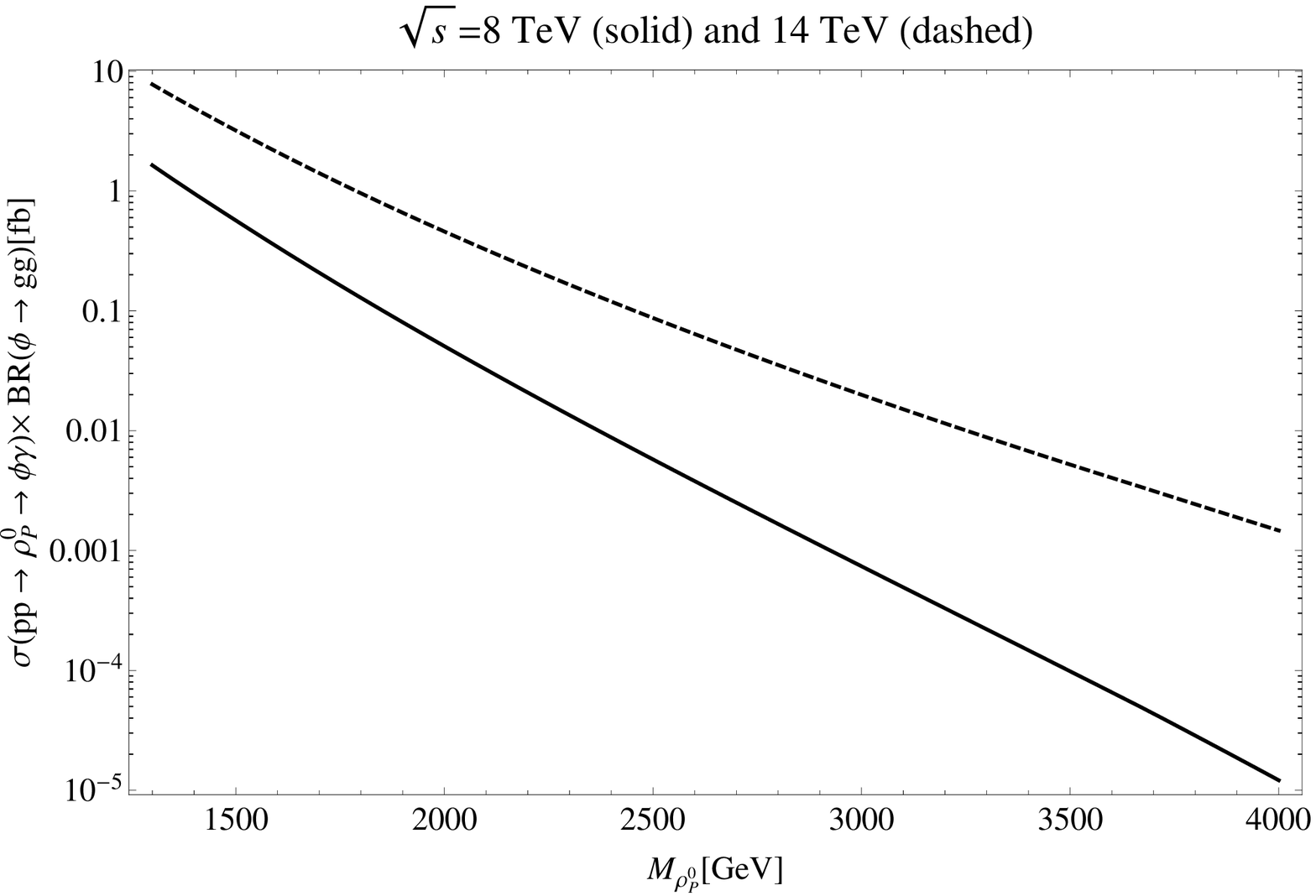}
\caption{{\bf Left panel:} $\sigma_{\rm DY}(pp \to \rho_{\theta}^0 \to \phi g) \times {\rm BR}(\phi \to gg)$ in the unit of fb as a function of $M_{\rho_{\theta}^0}$.  
{\bf Right panel:}  $\sigma_{\rm DY}(pp \to \rho_P^0 \to \phi \gamma) \times {\rm BR}(\phi \to gg)$ in the unit of fb as a function of $M_{\rho_P^0}$.
In both plots, solid and dashed curves correspond to cross sections for $\sqrt{s}=$ 8 and 14 TeV, respectively. Also, the branching ratio of the TD decaying into two gluons is taken to be ${\rm BR}(\phi \to gg) =75 \%$~\cite{Matsuzaki:2012mk}.  
\label{fig:signal}}
\end{center} 
\end{figure}

Since the TD dominantly decays to two gluons~\cite{Matsuzaki:2012mk}, 
we consider the process where the produced TD subsequently decays into two gluons. 
In Fig.~\ref{fig:signal}, we plot the cross sections of $pp \rightarrow \rho_{\theta}^0 \rightarrow \phi g \rightarrow g g g$ (left panel) and $pp \rightarrow \rho_P^0 \rightarrow \phi \gamma \rightarrow g g \gamma$ (right panel) as functions of each technirho mass. Here, we take 
${\rm BR}(\phi \to gg) =75 \%$, which can be read off from Ref.~\cite{Matsuzaki:2012mk}. 
The mass ranges (horizontal axes) of the plots are chosen in such a way that they are above the current LHC limit derived in the previous subsection and below the threshold of decay channels which involve the technipion(s). 
We can see that the cross section for the color-singlet channel (right panel in Fig.~\ref{fig:signal}) is rather small, and it may be challenging even at the 14 TeV LHC. We also estimated similar cross sections for an isotriplet technirho production followed by its decay into the TD and the electroweak gauge boson ($W/Z/\gamma$), and found that these cross sections are even smaller than that of the $\rho_P^0$ case. Meanwhile, the color-octet channel (left panel in Fig.~\ref{fig:signal}) has large cross sections, and can be a promising search channel at the 14 TeV LHC. A detailed collider study of this channel will be published elsewhere~\cite{KMTY}.

\section{Summary}
\label{sec:summary}
In this paper, 
we formulated a scale-invariant hidden local symmetry  
as a low-energy effective theory of walking technicolor,  
which includes the technidilaton, technipions, and technirho mesons as the low-lying spectra.  
As a benchmark for LHC phenomenology, 
our discussions have in particular focused on the one-family model of  walking technicolor 
with eight technifermion flavors, 
which can be -- at energy scales relevant to the reach of the LHC -- described by the scale-invariant hidden local symmetry 
based on the manifold 
$[SU(8)_L \times SU(8)_R]_{\rm global} \times SU(8)_{\rm local}/SU(8)_V$,  
where $SU(8)_{\rm local}$ is the hidden local symmetry and the global $SU(8)_L \times SU(8)_R$ symmetry is partially 
gauged by $SU(3) \times SU(2)_L \times U(1)_Y$ of the SM.     
Based on the scale-invariant hidden local symmetry, we evaluated the coupling properties of the technirho mesons 
and placed limits on the masses from the current LHC data. 
Then, implications for future LHC phenomenology were discussed by focusing on 
the technirho mesons produced through the Drell-Yan process.  
We found that the color-octet technirho decaying to the technidilaton along with the gluon 
is of interest as a discovery channel at the LHC, which would provide  a characteristic 
signature to probe the one-family model of walking technicolor.  
More detailed collider studies are in progress.

\acknowledgements
We would like to thank Koji Terashi for fruitful discussions.
This work was supported by 
the JSPS Grant-in-Aid for Scientific Research (S) \#22224003 and (C) \#23540300 (K.Y.). 

\appendix

\section{Interactions}
\label{app:int}

In this appendix, we summarize interactions that involve one technirho meson which can be obtained by expanding the $\chi^2 F_\sigma^2 {\rm tr}[\hat{\alpha}_{\mu||}^2]$ term in Eq.~(\ref{Lag}): 
\begin{eqnarray} 
\chi^2 F_\sigma^2 {\rm tr}[\hat{\alpha}_{\mu||}^2]
&=& F_\sigma^2 \left( 1 + \frac{2\phi}{F_\phi} + \cdots \right) 
\times \nonumber \\ 
&& 
{\rm tr} \left[ 
 ({\cal V}_\mu - V_\mu)^2 + \frac{i}{F_\pi^2} V_\mu [\partial^\mu\pi, \pi] + \frac{2i}{F_\pi} V_\mu [{\cal A}^\mu, \pi] 
 + \cdots
\right]
\,. 
\end{eqnarray}
Here, we focus on a set of  spectra, $(\rho_{\theta}^0, \rho^{\pm,3}_\Pi, \rho_P^0)$, which are expected to be 
produced through the Drell-Yan processes at the LHC. 
Using Eqs.(\ref{pi:para}), (\ref{rho:para}), and (\ref{v:a:para}), 
we thus derive the technirho couplings relevant  
for the LHC phenomenology.

 \subsection{$\rho-{\cal V}$ mixing terms}

\begin{eqnarray} 
{\cal L}_{{\cal V} \rho}  
&=& 
 - 2 g F_\sigma^2 {\rm tr}[  {\cal V}_\mu \rho^\mu ] 
\nonumber\\ 
&=&
 - 2 g F_\sigma^2
 \Bigg[ 
  \frac{g_s}{\sqrt{2}}  G_\mu^a \rho_{\theta}^{0 a \mu}
  + 
  e A_\mu \left\{
 \rho^{3 \mu}_{\Pi} + \frac{1}{\sqrt{3}} \rho^{0 \mu}_{P} 
 \right\}  
 \nonumber \\ 
&&
+ \frac{e}{2sc} Z_\mu  \left\{ 
  (c^2-s^2) \rho^{3 \mu}_\Pi - \frac{2}{\sqrt{3}} s^2 \rho^{0 \mu}_{P} 
 \right\} 
 + \frac{e}{2 s} 
 \left\{ 
  W_\mu^- \rho^{\mu+}_{\Pi} + {\rm H.c.} 
 \right\}
 \Bigg]
 \,, \label{v-rho:mixing}
\end{eqnarray}
where the charged rho-meson fields have been defined as 
\begin{equation} 
 \rho^{\mu \pm}_{\Pi} = \frac{\rho^{1 \mu}_{\Pi} \mp i \rho^{2 \mu}_{\Pi} }{\sqrt{2}}
\,. 
\end{equation}
Note the absence of $A-\rho_P^3, Z-\rho_P^3$, and $W^\pm-\rho_P^\mp$ terms due to 
a coincident cancellation of contributions 
between the techniquark and lepton sectors, which follows from the orthogonality of the $SU(8)_V$ generators.   
These terms are crucial for technirho meson productions through the Drell-Yan process at the LHC and 
allow the decays to the SM fermions by assuming vector-meson dominance.

\subsection{$\rho-f-f$ couplings}

The $\rho-{\cal V}$ terms in Eq.(\ref{v-rho:mixing}) 
allow the decays to the SM fermions by assuming vector-meson dominance via the 
couplings induced by the SM gauge-boson exchanges, evaluated at the $\rho$ on shell: 
\begin{eqnarray} 
{\cal L}_{\rho-f-f} 
&=& 
- 2 g F_\sigma^2 \Bigg[ 
\frac{ g_s^2}{\sqrt{2}} \frac{1}{ M_{\rho_{\theta}^0}^2} \bar{q} \gamma_\mu  \rho_{\theta a}^{\mu0 } \left( \frac{\lambda_a}{2} \right) q 
+ e \frac{1}{M_{\rho^3_\Pi}^2} J_\mu^{\rm em} \rho^{3\mu}_\Pi 
\nonumber \\ 
&& 
+ \frac{e(c^2-s^2)}{2 sc}\frac{1}{M_{\rho^3_\Pi}^2 - m_Z^2} J_\mu^Z \rho^{3\mu}_\Pi   
+ \frac{e}{2 s}\frac{1}{M_{\rho^\pm_\Pi}^2 - m_W^2} (J_\mu^{W^+} \rho^{+ \mu}_\Pi + {\rm H.c.}  )  
\nonumber \\ 
&& 
+ \frac{e}{\sqrt{3}} \frac{1}{M_{\rho_P^0}^2} J_\mu^{\rm em} \rho_P^{0\mu}  
- \frac{e s}{\sqrt{3} c} \frac{1}{M_{\rho_P^0}^2 - m_Z^2} J_\mu^{Z} \rho_P^{0\mu} 
\Bigg] 
\,, 
\end{eqnarray}
where  
\begin{eqnarray} 
  J_\mu^{\rm em} &=& e \sum_f \bar{f} \gamma_\mu Q_{\rm em}^f f 
  \,, \\ 
  J_\mu^Z &=& \frac{e}{sc} \sum_f \left[ \bar{f}_L \gamma_\mu (\tau^f_3 - s^2 Q_{\rm em}^f) f_L + \bar{f}_R \gamma_\mu (- s^2 Q_{\rm em}^f) f_R \right] 
\,, \\ 
 J_\mu^{W^\pm} &=& \frac{e}{\sqrt{2}s} \sum_f \bar{f}_L \gamma_\mu \tau^f_\pm f_L  
\,. 
\end{eqnarray}

 \subsection{$\rho - {\cal V}- \phi$ terms}

\begin{eqnarray} 
{\cal L}_{{\cal V} \rho \phi}  
&=& 
 - \frac{4 g F_\sigma^2}{F_\phi} \phi {\rm tr}[  {\cal V}_\mu \rho^\mu ] 
\nonumber\\ 
&=&
 - \frac{4 g F_\sigma^2}{F_\phi} \phi 
 \Bigg[ 
  \frac{g_s}{\sqrt{2}}  G_\mu^a \rho_{\theta}^{0 a \mu}
  + 
  e A_\mu \left\{
 \rho^{3 \mu}_{\Pi} + \frac{1}{\sqrt{3}} \rho^{0 \mu}_{P} 
 \right\}  
 \nonumber \\ 
&&
+ \frac{e}{2sc} Z_\mu  \left\{ 
  (c^2-s^2) \rho^{3 \mu}_\Pi - \frac{2}{\sqrt{3}} s^2 \rho^{0 \mu}_{P} 
 \right\} 
 + \frac{e}{2 s} 
 \left\{ 
  W_\mu^- \rho^{\mu+}_{\Pi} + {\rm H.c.} 
 \right\}
 \Bigg]
 \,. 
\end{eqnarray}
 Note again the absence of $A-\rho_P^3-\phi, Z-\rho_P^3-\phi$, and $W^\pm-\rho_P^\mp-\phi$ terms due to 
the orthogonality, namely, cancellation of contributions 
between the techniquark and lepton sectors.

 \subsection{$\rho-\pi-\pi$ terms}

The $\rho-\pi-\pi$ terms are decomposed into four parts:  
 \begin{eqnarray} 
  {\cal L}_{\rho -\pi -\pi} 
&=& 
2\, i\, g_{\rho\pi\pi} {\rm tr}[\rho_\mu [\partial^\mu \pi, \pi]] 
=
\frac{i g F_\sigma^2}{F_\pi^2} {\rm tr}[\rho_\mu [\partial^\mu \pi, \pi]] 
\nonumber \\ 
&=& 
\frac{i g F_\sigma^2}{F_\pi^2} {\rm tr}\Bigg[
\rho_{QQ}^\mu \left( \stackrel{\leftrightarrow}{\partial}_\mu \pi_{QQ}  \pi_{QQ} 
+ 
\stackrel{\leftrightarrow}{\partial}_\mu \pi_{QL}  \pi_{LQ}
\right)  
\nonumber\\ 
&& 
+ 
\rho_{QL}^\mu \left( \stackrel{\leftrightarrow}{\partial}_\mu \pi_{LQ}  \pi_{QQ}
+ 
\stackrel{\leftrightarrow}{\partial}_\mu \pi_{LL} \pi_{LQ}
\right)  
\nonumber\\ 
&& 
+ 
\rho_{LQ}^\mu \left( \stackrel{\leftrightarrow}{\partial}_\mu \pi_{QQ} \pi_{QL}
+ 
\stackrel{\leftrightarrow}{\partial}_\mu \pi_{QL} \pi_{LL} 
\right)  
\nonumber\\
&& 
+ 
\rho_{LL}^\mu \left( \stackrel{\leftrightarrow}{\partial}_\mu\pi_{LQ} \pi_{QL} 
+ 
\stackrel{\leftrightarrow}{\partial}_\mu \pi_{LL} \pi_{LL} 
\right)  
\Bigg] 
\nonumber\\ 
&=&
{\cal L}_{\rho -\pi-\pi} 
+ 
{\cal L}_{\rho_P-\pi-\pi} 
+ 
{\cal L}_{\rho_{T}-\pi-\pi} 
+ 
{\cal L}_{\rho_{\theta}-\pi-\pi} 
\,,  
\end{eqnarray}
where we have defined, for arbitrary fields $\bar{A}$ and $B$, 
\begin{eqnarray} 
\stackrel{\leftrightarrow}{\partial}_\mu A B \equiv \partial_\mu A B - A \partial_\mu B
\,. 
\end{eqnarray}

The $\rho_\Pi-\pi-\pi$ terms (arising from the $\rho_{QQ}$ and $\rho_{LL}$ terms) are 
\begin{eqnarray} 
{\cal L}_{\rho_\Pi-\pi-\pi} 
&=& 
 \frac{i g F_\sigma^2}{F_\pi^2} 
 \Bigg[
 \frac{1}{4} i \epsilon^{ijk} \partial_\mu \Pi^i \Pi^j \rho^{k \mu}_{\Pi} 
+ \frac{1}{4} i \epsilon^{ijk} \stackrel{\leftrightarrow}{\partial}_\mu \Pi^i P^j \rho^{k \mu}_{\Pi} 
\nonumber \\ 
&& 
+ 
\frac{1}{4} i \epsilon^{ijk} \partial_\mu P^i P^j \rho^{k \mu}_{\Pi} 
+ 
\frac{1}{4} i \epsilon^{ijk} \stackrel{\leftrightarrow}{\partial}_\mu \bar{T}^i T^j \rho^{k \mu}_{\Pi} 
\Bigg] 
\,. 
 \end{eqnarray}
 Note the absence of $\rho^i_\Pi-T^i-T^0$ terms due to the accidental cancellation between the techniquark and technilepton sectors. 
The $\rho_P-\pi-\pi$ terms (arising from the $\rho_{QQ}$ and $\rho_{LL}$ terms) are 
\begin{eqnarray} 
 {\cal L}_{\rho_P-\pi-\pi} 
 &=&   
 \frac{ig F_\sigma^2}{F_\pi^2} \Bigg[
 - \frac{1}{2 \sqrt{3}} i \epsilon^{ijk} \partial_\mu P^i P^j \rho_P^{k \mu} 
 - \frac{1}{4 \sqrt{3}} i \epsilon^{ijk} \stackrel{\leftrightarrow}{\partial}_\mu \bar{T}^i T^j \rho_P^{k \mu} 
\nonumber \\  
&&
  - \frac{1}{2 \sqrt{3}} \left( \stackrel{\leftrightarrow}{\partial}_\mu \bar{T}^i T^0 + \stackrel{\leftrightarrow}{\partial}_\mu \bar{T}^0T^i \right) 
\rho^{i \mu}_{P}
- \frac{1}{2 \sqrt{3}} \left( \stackrel{\leftrightarrow}{\partial}_\mu \bar{T}^i T^i + 
\stackrel{\leftrightarrow}{\partial}_\mu \bar{T}^0 T^0 \right) \rho_P^{0\mu} 
 \Bigg]
 \,. 
\end{eqnarray}
Note that there is no $\rho_P-\Pi-\Pi$ term due to an accidental 
cancellation between the techniquark and technilepton sectors.

The $\rho_{T}-\pi-\pi$ terms (arising from the $\rho_{QL}$ and $\rho_{LQ}$ terms) are expressed as
 \begin{eqnarray}
{\cal L}_{\rho_{T}-\pi-\pi} 
&=& 
 \frac{ig F_\sigma^2}{F_\pi^2} \Bigg[
\frac{i}{2 \sqrt{2}} \epsilon^{ijk} \stackrel{\leftrightarrow}{\partial}_\mu \bar{T}^i \theta_a^j \left( \frac{\lambda_a}{2} \right) \rho_{T}^{k \mu} 
+ 
\frac{1}{2\sqrt{2}} \stackrel{\leftrightarrow}{\partial}_\mu \bar{T}^0 \theta_a^i \left( \frac{\lambda_a}{2} \right) \rho_{T}^{i \mu} 
+
\frac{1}{2\sqrt{2}} \stackrel{\leftrightarrow}{\partial}_\mu \bar{T}^i \theta_a^0 \left(  \frac{\lambda_a}{2} \right) \rho_{T}^{i \mu} 
\nonumber\\ 
&& 
+
\frac{i}{2\sqrt{3}} \epsilon^{ijk} \stackrel{\leftrightarrow}{\partial}_\mu \bar{T}^i P^j \rho_{T}^{k \mu} 
+ 
\frac{1}{2\sqrt{3}} \stackrel{\leftrightarrow}{\partial}_\mu \bar{T}^0 P^i \rho_{T}^{i \mu} 
+
\frac{1}{2\sqrt{3}} \stackrel{\leftrightarrow}{\partial}_\mu \bar{T}^i P^0 \rho_{T}^{i \mu} 
\nonumber \\ 
&& 
+ 
\frac{1}{2\sqrt{2}} \stackrel{\leftrightarrow}{\partial}_\mu \bar{T}^i  \theta_a^i \left( \frac{\lambda_a}{2} \right) \rho_{T}^{0\mu} 
+ 
\frac{1}{2\sqrt{2}} \stackrel{\leftrightarrow}{\partial}_\mu \bar{T}^0  \theta_a^0 \left( \frac{\lambda_a}{2} \right) \rho_{T}^{0\mu} 
\nonumber \\ 
&&
+ 
\frac{1}{2\sqrt{3}} \stackrel{\leftrightarrow}{\partial}_\mu \bar{T}^i P^i \rho_{T}^{0\mu} 
+
\frac{1}{2\sqrt{3}} \stackrel{\leftrightarrow}{\partial}_\mu \bar{T}^0 P^0 \rho_{T}^{0\mu} 
\Bigg]
\nonumber\\ 
&& 
+ 
{\rm H.c.}  
\,.  
\end{eqnarray} 
The $\rho_{\theta}-\pi-\pi$ terms (arising from the $\rho_{QQ}$ term) are expressed as 
\begin{eqnarray} 
{\cal L}_{\rho_{\theta}-\pi-\pi} 
&=& 
 \frac{ig F_\sigma^2}{F_\pi^2} \Bigg[
- \frac{1}{\sqrt{2}} f^{abc} \epsilon^{ijk} \partial_\mu \theta_a^i \theta_b^j \rho_{\theta c}^{k \mu}  
- \frac{1}{2\sqrt{2}} \epsilon^{ijk} \stackrel{\leftrightarrow}{\partial}_\mu \bar{T}^j  \left( \frac{\lambda_a}{2} \right) T^i \rho_{\theta a}^{k\mu}
\nonumber \\ 
&& 
- \frac{1}{2\sqrt{2}} \stackrel{\leftrightarrow}{\partial}_\mu \bar{T}^i  \left( \frac{\lambda_a}{2} \right) T^0  \rho_{\theta a}^{i\mu} 
- \frac{1}{2\sqrt{2}} \stackrel{\leftrightarrow}{\partial}_\mu \bar{T}^0  \left( \frac{\lambda_a}{2} \right) T^i  \rho_{\theta a}^{i\mu}
\nonumber\\ 
&& 
- \frac{1}{2\sqrt{2}}  \stackrel{\leftrightarrow}{\partial}_\mu \bar{T}^i  \left( \frac{\lambda_a}{2} \right) T^i \rho_{\theta a}^{0\mu}
 - \frac{1}{2\sqrt{2}}  \stackrel{\leftrightarrow}{\partial}_\mu \bar{T}^0  \left( \frac{\lambda_a}{2} \right) T^0 \rho_{\theta a}^{0\mu}
\Bigg] 
\,. 
\end{eqnarray}

 \subsection{$\rho - {\cal A} - \pi$ terms}

 The $\rho-{\cal A}-\pi$ terms are constructed from three parts: 
 \begin{eqnarray} 
  {\cal L}_{\rho-{\cal A}-\pi} 
&=& 
\frac{2i g F_\sigma^2}{F_\pi} {\rm tr}[ \rho_\mu [{\cal A}^\mu, \pi] ] 
\nonumber \\ 
&=& 
{\cal L}_{\rho-{\cal A}-\pi} 
+ 
{\cal L}_{\rho_P-{\cal A}-\pi} 
+ 
{\cal L}_{\rho_{T}-{\cal A}-\pi} 
\,. 
 \end{eqnarray}

The $\rho_\Pi-{\cal A}-\pi$ terms are  
\begin{equation} 
{\cal L}_{\rho_\Pi-{\cal A}-\pi} 
=
 \frac{2i g F_\sigma^2}{F_\pi} 
 \Bigg[ 
 \frac{e}{4sc} Z_\mu \left( \rho^{+\mu}_\Pi \Pi^- - \rho^{-\mu}_\Pi \Pi^+  \right) 
+ 
\frac{e}{4s} \left[ W_\mu^+ \left(  \rho^{- \mu}_\Pi \Pi^3 - \rho^{3\mu}_\Pi \Pi^- \right) + {\rm H.c.}  \right]
\Bigg] 
\end{equation}

The $\rho_P-{\cal A}-\pi$ terms are  
\begin{eqnarray} 
{\cal L}_{\rho_P-{\cal A}-\pi} 
&=&
 \frac{2i g F_\sigma^2}{F_\pi} 
 \Bigg[ 
\frac{e}{4sc} Z_\mu \left( \rho_P^{+\mu} P^-  - \rho_P^{-\mu} P^+ \right)  
 + 
 \frac{e}{4s} \left\{ W_\mu^+ \left( \rho_P^{-\mu} P^3  -  \rho_P^{+\mu} P^3  \right)
 + 
W_\mu^- \left( \rho_P^{3 \mu} P^+  -  \rho_P^{+\mu} P^-  \right) \right\} 
 \Bigg] 
 \,, 
\end{eqnarray}
where 
\begin{equation} 
\rho_P^{\pm \mu} 
\equiv 
\frac{\rho_P^{1 \mu} \mp i \rho_P^{2 \mu}}{\sqrt{2}}
\,. 
\end{equation}

The $\rho_{T}-{\cal A}-\pi$ terms are 
\begin{eqnarray} 
{\cal L}_{\rho_{T}-{\cal A}-\pi} 
&=&  
 \frac{2i g F_\sigma^2}{F_\pi} 
 \Bigg[ 
  \frac{e}{4sc} Z_\mu \left( \bar{T}^- \rho^{+ \mu}_T - \bar{T}^+ \rho^{- \mu}_T \right) 
  + 
  \frac{e}{4s} \left\{ W_\mu^+ \left( \bar{T}^3 \rho_{T}^{- \mu} - \bar{T}^- \rho_{T}^{3 \mu} \right) 
+ W_\mu^- \left( \bar{T}^+ \rho_{T}^{3 \mu} - \bar{T}^3 \rho_{T}^{+ \mu} \right) 
  \right\} 
 \Bigg]
\nonumber \\ 
&&  
 + {\rm H.c.} 
 \,, 
\end{eqnarray}
where 
\begin{eqnarray} 
 T^\pm &\equiv& \frac{T^1 \mp i T^2}{\sqrt{2}}
\,, \qquad 
\bar{T}^\pm = (T^\pm)^\dag 
\,\\ 
\rho_{T}^{\pm \mu}   &\equiv& 
\frac{\rho_{T}^{1 \mu} \mp i \rho_{T}^{2 \mu}}{\sqrt{2}}
\,. 
\end{eqnarray}

\subsection{$\rho-{\cal A}-{\cal V}$ terms} 

It turns out that all the terms involving the $\rho_{\theta}^0, \rho^{\pm, 3}_\Pi$ and $\rho_P^0$ fields vanish 
due to the $SU(8)_V$ symmetry, so that 
\begin{equation} 
  {\cal L}_{V{\cal A}{\cal V}} \Bigg|_{\rho_{\theta}^0, \rho^{\pm, 3}_\Pi, \rho_P^0} = 0
\,. 
\end{equation}

\section{Partial decay widths of the technirho mesons} 
\label{app:Pdecay}

In this appendix, we summarize the partial decay rates of the technirho mesons studied in this paper ($\rho_{\theta}^0$, $\rho_P^0$, and $\rho^{\pm ,3}_\Pi$) that are relevant for collider phenomenology.

\subsection{The $\rho_{\theta}^0$ partial decay rates} 

\begin{eqnarray}  
\Gamma(\rho_{\theta}^0 \to \bar{T}^0 T^0) 
&=& 
\frac{1}{96\pi} \left(  \frac{g F_\sigma^2}{2 \sqrt{2} F_\pi^2}  \right)^2 \frac{\left[ M_{\rho_{\theta}^0}^2 - 4 M_{T^0}^2 \right]^{3/2}}{M_{\rho_{\theta}^0}^2}  
\,, \\
\Gamma(\rho_{\theta}^0 \to \bar{q}q) 
&=& 
\frac{1}{24 \pi} \left(  \frac{ \sqrt{2} g_s^2 g F_\sigma^2}{M_{\rho_{\theta}^0}^2}  \right)^2  \left( \frac{M_{\rho_{\theta}^0}^2 + 2 m_q^2}{M_{\rho_{\theta}^0}^2} \right)
\left[ M_{\rho_{\theta}^0}^2 - 4 m_q^2 \right]^{1/2}  
\,, \\
\Gamma(\rho_{\theta}^0 \to g \phi) 
&=& 
\frac{1}{8 \pi} \left(  \frac{2 \sqrt{2} g_s g F_\sigma^2}{F_\phi}  \right)^2 \left( \frac{M_{\rho_{\theta}^0}^2 - M_\phi^2}{M_{\rho_{\theta}^0}^3} \right)
\,. 
\end{eqnarray}

\subsection{The $\rho_P^0$ partial decay rates} 

\begin{eqnarray} 
\Gamma(\rho_P^0 \to \bar{T}^i T^i) 
&=& 
\frac{3}{16 \pi} \left( \frac{g F_\sigma^2}{2 \sqrt{3} F_\pi^2} \right)^2 
\frac{\left[ M_{\rho_P^0}^2 - 4 M_{T^i}^2  \right]^{3/2}}{M_{\rho_P^0}^2} 
\,, \\ 
\Gamma(\rho_P^0 \to \bar{T}^0 T^0) 
&=& 
\frac{1}{16 \pi} \left( \frac{g F_\sigma^2}{2 \sqrt{3} F_\pi^2} \right)^2 
\frac{\left[ M_{\rho_P^0}^2 - 4 M_{T^0}^2  \right]^{3/2}}{M_{\rho_P^0}^2} 
\,, 
\end{eqnarray} 
\begin{eqnarray} 
\Gamma(\rho_P^0 \to f \bar{f}) 
&=& 
\frac{N_c^{(f)}}{12 \pi} 
 \left( \frac{2 e^2 g F_\sigma^2}{\sqrt{3}}  \right)^2  
\left[ \frac{[G_V^{\rho_P^0}]^2 (M_{\rho_P^0}^2 + 2 m_f^2) + [G_A^{\rho_P^0}]^2 (M_{\rho_P^0}^2 - 4 m_f^2)}{M_{\rho_P^0}^2}  \right]
 \sqrt{M_{\rho_P^0}^2 - 4 m_f^2} 
\,,\\ 
\Gamma(\rho_P^0 \to \phi \gamma) 
&=& 
\frac{1}{16 \pi} \left( \frac{4 e g F_\sigma^2}{\sqrt{3} F_\phi}  \right)^2 
\left(\frac{M_{\rho_P^0}^2 - M_\phi^2}{M_{\rho_P^0}^3}  \right) 
\,, \label{rhoP-phigamma} \\ 
\Gamma(\rho_P^0 \to \phi Z) 
&=& 
\frac{1}{16 \pi} \left( \frac{4 e sg F_\sigma^2}{\sqrt{3} c F_\phi}  \right)^2 
  \frac{\sqrt{ \left( M_{\rho_P^0}^2 - (M_\phi + m_Z)^2 \right) \left(  M_{\rho_P^0}^2 - (M_\phi - m_Z)^2 \right)}}{M_{\rho_P^0}^3} 
\,, 
\end{eqnarray}
where $N_c^{(f)}=1(3)$ for leptons (quarks) and 
\begin{eqnarray} 
G_V^{\rho_P^0}&=&
\frac{Q_{\rm em}^f}{M_{\rho_P^0}^2} - \frac{\tau_3^f - 2 s^2 Q_{\rm em}^f }{2c^2 (M_{\rho_P^0}^2 - m_Z^2)} 
\,, \\ 
G_A^{\rho_P^0} 
&=& 
\frac{\tau_f^3}{2c^2 (M_{\rho_P^0}^2 - m_Z^2)} 
\,. 
\end{eqnarray}

\subsection{The $\rho^3_\Pi$ partial decay rates}

\begin{eqnarray} 
\Gamma(\rho^3_\Pi \to W_L W_L) 
&=& 
\frac{1}{48 \pi} \left( \frac{g F_\sigma^2}{4 F_\pi^2} \right)^2 
\frac{\left[ M_{\rho^3_\Pi}^2 - 4 m_W^2 \right]^{3/2}}{M_{\rho^3_\Pi}^2}
\,, \\ 
\Gamma(\rho^3_\Pi \to P^+ P^-) 
&=& 
\frac{1}{48 \pi} \left( \frac{g F_\sigma^2}{4 F_\pi^2} \right)^2 
\frac{\left[ M_{\rho^3_\Pi}^2 - 4 M_{P^\pm_\Pi}^2 \right]^{3/2}}{M_{\rho^3_\Pi}^2}
\,, \\ 
\Gamma(\rho^3_\Pi \to \bar{T}^\pm T^\mp) 
&=& 
\frac{1}{8 \pi} \left( \frac{g F_\sigma^2}{4 F_\pi^2} \right)^2 
\frac{\left[ M_{\rho^3_\Pi}^2 - 4 M_{T^\pm}^2 \right]^{3/2}}{M_{\rho^3_\Pi}^2}
\,, \\ 
\Gamma(\rho^3_\Pi \to W_L^\pm P^\mp) 
&=& 
\frac{1}{24\pi} \left( \frac{g F_\sigma^2}{4 F_\pi^2}  \right)^2 
\frac{\left[ ( M_{\rho^3_\Pi}^2 - (m_W + M_{P^\pm})^2 )(M_{\rho^3_\Pi}^2 - (m_W - M_{P^\pm})^2)  \right]^{3/2}}{M_{\rho^3_\Pi}^5}
\,, \\ 
\Gamma(\rho^3_\Pi \to f\bar{f}) 
&=& 
\frac{N_c^{(f)}}{12 \pi} 
(2 e^2 g F_\sigma^2 )^2  
\left[ \frac{[G_V^{\rho^3_\Pi}]^2 (M_{\rho^3_\Pi}^2 + 2 m_f^2) + [G_A^{\rho^3_\Pi}]^2 (M_{\rho^3_\Pi}^2 - 4 m_f^2)}{M_{\rho^3_\Pi}^2}  \right]
\sqrt{M_{\rho^3_\Pi}^2 - 4 m_f^2}
\,, \nonumber \\ 
\Gamma(\rho^3_\Pi \to \phi \gamma) 
&=& 
\frac{1}{16 \pi} \left( \frac{4 e g F_\sigma^2}{F_\phi} \right)^2 \left( \frac{M_{\rho^3_\Pi}^2 - M_\phi^2}{M_{\rho^3_\Pi}^3} \right)
\,, \\ 
\Gamma(\rho^3_\Pi \to \phi Z) 
&=&
\frac{1}{16 \pi} \left( \frac{2 e (c^2- s^2) g F_\sigma^2}{sc F_\phi} \right)^2 
\frac{\sqrt{ ( M_{\rho^3_\Pi}^2 - (m_Z + M_{\phi})^2 )(M_{\rho^3_\Pi}^2 - (m_Z - M_{\phi})^2)  }}{M_{\rho^3_\Pi}^3}
\,, 
\end{eqnarray}
where $W_L^\pm \equiv \Pi^\pm$ and 
\begin{eqnarray} 
G_V^{\rho^3_\Pi}&=&
\frac{Q_{\rm em}^f}{M_{\rho^3_\Pi}^2} + \frac{c^2-s^2}{s^2 c^2} \frac{\tau_3^f - 2 s^2 Q_{\rm em}^f }{4  (M_{\rho_\Pi^3}^2 - m_Z^2)} 
\,, \\ 
G_A^{\rho^3_\Pi} 
&=& 
- \frac{c^2-s^2}{s^2 c^2 } \frac{\tau_f^3}{4 (M_{\rho^3_\Pi}^2 - m_Z^2)} 
\,. 
\end{eqnarray}

\subsection{The $\rho^\pm_\Pi$ partial decay rates}

\begin{eqnarray} 
\Gamma(\rho^\pm_\Pi \to W_L^\pm Z_L) 
&=& 
\frac{1}{48 \pi} \left( \frac{g F_\sigma^2}{4 F_\pi^2} \right)^2 
\frac{\left[ (M_{\rho^\pm_\Pi}^2 - (m_W+m_Z)^2)(M_{\rho^\pm_\Pi}^2 - (m_W-m_Z)^2)\right]^{3/2}}{M_{\rho^\pm_\Pi}^5}
\,, \\ 
\Gamma(\rho^\pm_\Pi \to P^\pm P^3) 
&=& 
\frac{1}{48 \pi} \left( \frac{g F_\sigma^2}{4 F_\pi^2} \right)^2 
\frac{\left[ (M_{\rho^\pm_\Pi}^2 - (M_{P^\pm}+M_{P^3})^2)(M_{\rho^\pm_\Pi}^2 - (M_{P^\pm}-M_{P^3})^2)\right]^{3/2}}{M_{\rho^\pm_\Pi}^5}
\,, \\ 
\Gamma(\rho^\pm_\Pi \to \bar{T}^\pm T^3) 
&=& \Gamma(\rho^\pm_\Pi \to \bar{T}^3 T^\pm) 
\nonumber\\ 
&=& 
\frac{1}{8 \pi} \left( \frac{g F_\sigma^2}{4 F_\pi^2} \right)^2 
\frac{\left[ (M_{\rho^\pm_\Pi}^2 - (M_{T^\pm}+M_{T^3})^2)(M_{\rho^\pm_\Pi}^2 - (M_{T^\pm}-M_{T^3})^2)\right]^{3/2}}{M_{\rho^\pm_\Pi}^5}
\,, \\ 
\Gamma(\rho^\pm_\Pi \to W_L^\pm P^3) 
&=& 
\frac{1}{24\pi} \left( \frac{g F_\sigma^2}{4 F_\pi^2}  \right)^2 
\frac{\left[ (M_{\rho^\pm_\Pi}^2 - (m_W+M_{P^3})^2)( M_{\rho^\pm_\Pi}^2 - (m_W-M_{P^3})^2) \right]^{3/2}}{M_{\rho^\pm_\Pi}^5}
\,, \\ 
\Gamma(\rho^\pm_\Pi \to Z_L P^\pm) 
&=& 
\frac{1}{24\pi} \left( \frac{g F_\sigma^2}{4 F_\pi^2}  \right)^2 
\frac{\left[ (M_{\rho^\pm_\Pi}^2 - (m_Z+M_{P^\pm})^2)(M_{\rho^\pm_\Pi}^2 - (m_Z-M_{P^\pm})^2) \right]^{3/2}}{M_{\rho^\pm_\Pi}^5}
\,, \\ 
\Gamma(\rho^\pm_\Pi \to f_1 \bar{f}_2) 
&=& 
\frac{N_c^{(f)}}{48 \pi} 
\left( 
 \frac{e^2 g F_\sigma^2}{\sqrt{2} s^2 (M_{\rho^\pm_\Pi}^2 - m_W^2) }  
\right)^2 
\frac{\left( 2 M_{\rho^\pm_\Pi}^4 - (m_{f_1}^2 +  m_{f_2}^2) M_{\rho^\pm_\Pi}^2 - (m_{f_1}^2 - m_{f_2}^2)^2 \right)}{M_{\rho^\pm_\Pi}^4} 
\nonumber \\ 
&&
\times  
\frac{\sqrt{ (M_{\rho^\pm_\Pi}^2 - (m_{f_1}+m_{f_2})^2)( M_{\rho^\pm_\Pi}^2 - (m_{f_1}-m_{f_2})^2)}}{M_{\rho^\pm_\Pi}}
\,,\\ 
\Gamma(\rho^\pm_\Pi \to \phi W^\pm) 
&=&
\frac{1}{16 \pi} \left( \frac{2 e g F_\sigma^2}{s F_\phi} \right)^2 
\frac{\sqrt{ ( M_{\rho^\pm_\Pi}^2 - (m_W + M_{\phi})^2 )(M_{\rho^\pm_\Pi}^2 - (m_W - M_{\phi})^2)  }}{M_{\rho^\pm_\Pi}^3}
\,, 
\end{eqnarray}
where $Z_L \equiv \Pi^3$.


\end{document}